\documentclass[12pt]{article}

\usepackage{amsmath}
\usepackage{graphicx,psfrag,epsf}
\usepackage{enumerate}
\usepackage{natbib}
\usepackage{url}
\usepackage{bbm}
\usepackage{amsfonts,url}
\usepackage[normalem]{ulem}
\usepackage{amssymb}
\usepackage{amsmath}
\usepackage{amstext,verbatim,graphicx,ifthen,multirow,amsthm}
\usepackage{bm}
\usepackage{natbib}
\usepackage[bf]{caption}
\usepackage[nodisplayskipstretch]{setspace}
\usepackage[breaklinks=true,hidelinks]{hyperref}
\hypersetup{
    colorlinks,
    linkcolor={red!50!black},
    citecolor={blue!50!black},
    urlcolor={blue!80!black}
}
\usepackage{breakcites}
\usepackage{mathtools}
\usepackage{booktabs}
\usepackage{threeparttable}
\usepackage{tikz}
\usetikzlibrary{arrows.meta}
\usepackage{xfrac}
\usepackage{subcaption}
 \usepackage{color-edits}
 \addauthor{js}{blue}
 \addauthor{mw}{green}
 \addauthor{lb}{red}
 \addauthor{gwi}{orange}
\usepackage{xcolor}

\renewcommand{\mathbf}{\boldsymbol}
\newcommand{\indep}{\perp\!\!\!\perp}

\usepackage{siunitx}
\usepackage{rotating}

\newcommand{\pcb}{{\sc PCB}}
\newcommand{\bias}{\ensuremath{\mathrm{Biased}}}
\newcommand{\byn}{\mathbf{Y}(0)}
\newcommand{\bye}{\mathbf{Y}(1)}
\newcommand{\by}{\mathbf{Y}}
\newcommand{\bw}{\mathbf{W}}

\newcommand{\bu}{\mathbf{U}}
\newcommand{\bv}{\mathbf{V}}
\newcommand{\buu}{\mathbf{u}}
\newcommand{\bvv}{\mathbf{v}}
\newcommand{\ccc}{\mathrm{C}}

\newcommand{\calm}{{\cal M}}
\newcommand{\calmi}{{\cal M}^{(i)}}

\newcommand{\ddd}{\mathrm{DiM}}

\newcommand{\gsc}{\mathrm{GSC}}
\newcommand{\hmmv}{\hat{\mmv}}
\newcommand{\hhh}{\mathrm{H}}
\newcommand{\musc}{\mathrm{MUSC}}
\newcommand{\msc}{\mathrm{MSC}}
\newcommand{\momega}{\mathbf{M}}

\newcommand{\momegai}{\mathbf{M}^{(i)}}

\newcommand{\mme}{\mathbb{E}}
\newcommand{\mmv}{\mathbb{V}}
\newcommand{\nttt}{N_{\rm T}}
\newcommand{\nccc}{N_{\rm C}}

\newcommand{\ssc}{\mathrm{SC}}
\newcommand{\sumi}{\sum_{i=1}^N}
\newcommand{\sumj}{\sum_{j=1}^N}

\newcommand{\sumt}{\sum_{t=1}^T}

\newcommand{\sumit}{\sum_{i=1}^N \sum_{t=1}^T}
\newcommand{\sumituv}{\sum_{i=1}^N \sum_{t=1}^T U_i V_t}

\newcommand{\taut}{\tau}
\newcommand{\taup}{\tau^\mathrm{POP}}
\newcommand{\tauv}{\tau^{\vvv}}
\newcommand{\tauh}{\tau^{\hhh}}
\newcommand{\tbu}{\tilde{\bu}}

\newcommand{\usc}{\mathrm{USC}}
\newcommand{\unbias}{$\mathrm{Unbiased}$}
\newcommand{\vvv}{\mathrm{V}}
\newcommand{\wit}{W_{it}}
\newcommand{\yit}{Y_{it}}

\newcommand{\yite}{Y_{it}(1)}
\newcommand{\yitn}{Y_{it}(0)}

\newtheorem{assumption}{Assumption}

\newtheorem{example}{Example}

\newtheorem{lemma}{Lemma}
\newtheorem{prop}{Proposition}

\def\monthname{\ifcase\month\or
  January\or February\or March\or April\or May\or June\or July\or
  August\or September\or October\or November\or December\fi}
\numberwithin{equation}{section}

\DeclareMathOperator*{\argmin}{arg\,min}

\let\oldappendix\appendix
\renewcommand{\appendix}{
\oldappendix
\small\parindent 0cm\setcounter{equation}{0}
\renewcommand{\theequation}{\thesection.\arabic{equation}}
\setcounter{lemma}{0}\renewcommand{\thelemma}{\thesection.\arabic{lemma}}
\setcounter{theorem}{0}\renewcommand{\thetheorem}{\thesection.\arabic{theorem}}
\setcounter{example}{0}\renewcommand{\theexample}{\thesection.\arabic{example}}
}

  \title{\bf A Design-Based Perspective\\ on Synthetic Control Methods}
  \author{
  Lea Bottmer\hspace{.2cm}\\
    Department of Economics, Stanford University
    \and
    Guido W. Imbens \\
    Graduate School of Business \\ and Department of Economics, Stanford University
    \and
    Jann Spiess\\
    Graduate School of Business, Stanford University
    \and
    Merrill Warnick\\
    Department of Economics, Stanford University
    }
    \date{July 12, 2023}
\begin{document}

\def\spacingset#1{\renewcommand{\baselinestretch}
{#1}\small\normalsize}
\spacingset{1}

 \maketitle

\bigskip
\begin{abstract}
Since their introduction in \citet{abadie2003}, Synthetic Control (SC) methods have quickly become one of the leading methods for estimating causal effects in observational studies in settings with panel data. Formal discussions often motivate SC methods  by the assumption that the potential outcomes were generated by a factor model.
Here we study SC methods from a design-based perspective, assuming a model for the selection of the  treated unit(s) and period(s). We show that the standard SC estimator is generally biased under random assignment.  We propose a Modified Unbiased Synthetic Control (MUSC) estimator that guarantees unbiasedness under random assignment and derive its exact, randomization-based, finite-sample variance. We also propose an unbiased estimator for this variance. We document in settings with real data that under random assignment, SC-type estimators can have root mean-squared errors  that are substantially lower than that of
other common estimators.  We show that such an improvement is weakly guaranteed if the treated period is similar to the other periods, for example, if the treated period was randomly selected.
While our results only directly apply in settings where treatment is assigned randomly, we believe that they can complement model-based approaches even for observational studies.
\end{abstract}

\noindent%
\vfill

\newpage
\spacingset{1.8} %

\setcounter{section}{0}
\section{INTRODUCTION}

Synthetic Control (SC) methods for estimating causal effects have become popular in empirical work in the social sciences since their introduction 
in \citet*{abadie2003, abadie2010synthetic, abadie2014}. 
Typically, the properties of the  SC estimator are studied under model-based assumptions about the distribution of the potential outcomes in the absence of the intervention, often assuming the potential outcomes follow a factor model with noise. 
Here we take a design-based approach to Synthetic Control methods where   we  make assumptions about the assignment  of the unit/time-period pairs to treatment, and consider properties of the estimators conditional on the potential outcomes. 
We find that  in this setting, the original SC estimator is generally biased. We propose a modification of the SC estimator,  labeled the
 Modified Unbiased Synthetic Control (MUSC) estimator, 
which is unbiased under random assignment of the treatment, derive the exact variance of this estimator, and propose an unbiased estimator for this variance.

Studying the properties of SC-type estimators under design-based assumptions serves two distinct purposes. 
First, it suggests an important role for SC methods in the analysis of  experimental data 
and second, it leads to new insights into the properties of  SC methods in observational studies. We show that in experimental settings
SC-type methods (including both the original SC estimator and our proposed MUSC estimator) can have substantially better root-mean-squared-error (RMSE) properties than the standard difference-in-means (DiM) estimator, with this improvement guaranteed  under time randomization and a large number of time periods.
Beyond the analysis of existing data, our design-based results can be relevant for choosing  assignment probabilities when planning a randomized experiment, in the spirit of Rubin's adage that ``design trumps analysis'' \citep{rubin2008objective}.
Our approach complements  that in \citet{abadie2021synthetic}, who analyse the choice of treated units in a non-randomized setting to optimize precision. Instead, we focus on the  implications of randomizing units to assignment for inference.
To illustrate the  benefits of the MUSC estimator in experimental settings, we simulate an experiment based on average log wage data observed across 10 states over 40 years. We randomly select one state to be treated in the last period, and compare $(i)$ the DiM (difference-in-means) estimator, $(ii)$ the standard SC estimator, $(iii)$ our proposed MUSC estimator, and $(iv)$ the widely used DiD (difference-in-differences) estimator.
As  reported in \autoref{table_illustration}, the SC estimator is biased. Although in this example the bias of the SC estimator is modest, the bias can be arbitrarily large. We also find that the RMSE is substantially lower for the SC and MUSC estimators relative to the RMSE of the DiM estimator  because the SC and MUSC estimators effectively use the information in the pre-treatment periods. Finally, we find that the proposed variance estimator is accurate in this setting for all four estimators.

\begin{table}[!htbp] \centering 
  \caption{Simulation Experiment Based on CPS Average Log Wage by State and Year} 
  \label{table_illustration} 
\begin{tabular}{l S[table-format=3.2] S[table-format=3.2] S[table-format=3.2] S[table-format=3.2] S[table-format=3.2]}
    \toprule
  & {DiM}     & {SC} & {MUSC}   & {DiD}  \\ \midrule
Bias & 0 & -0.007 & 0 & 0\\
RMSE & 0.105 & 0.051 & 0.048 &  0.060 \\
Average standard error & 0.105 & 0.051 & 0.048  & 0.060\\
\bottomrule
\end{tabular}\\ \vskip0.2cm
{\footnotesize DiM: Difference in Means estimator; SC: Synthetic Control estimator of
\cite{abadie2010synthetic}; MUSC: Modified Unbiased Synthetic Control estimator; DiD: Difference-in-Differences estimator.}
\end{table}

The second contribution of the article concerns insights into Synthetic Control methods in observational studies.
Inference for Synthetic Control estimators has proven to be a challenge in many applications. Part of this reflects the difficulty in specifying the data generating process. As Manski and Pepper write in the context of a similar setting with data from the fifty US states:
``[M]easurement of
statistical precision requires specification of a sampling process
that generates the data. Yet we are unsure what type of
sampling process would be reasonable to assume in this
application. One would have to view the existing United
States as the sampling realization of a random process
defined on a superpopulation of alternative nations.'' \citep[][p. 234]{manski2018right}. We share the concerns raised by Manski and Pepper.
When the sample at hand can be viewed as a random sample from a well-defined population, it is natural and common to use sampling-based standard errors. When the causal variables of interest can be viewed as randomly assigned it is natural to use design-based standard errors, irrespective of the origin of the sample. When neither applies, and researchers still wish to report measures of uncertainty, they face choices about viewing outcomes or treatments as stochastic. 
This is the case in many Synthetic Control applications. There is a fixed set of units, {\it e.g.,} the fifty states of the US, clearly not sampled randomly from a well-defined population, with a given set of regulations, clearly not randomly assigned. 
Much of the Synthetic Control literature has chosen to focus on viewing outcomes as random in a way that motivates using sampling-based standard errors. In contrast here we analyze the  data as if the treatments are stochastic and propose design-based standard errors. 
Although others may disagree, we view this still as a natural starting point for many causal analyses, possibly after some adjustment for observed covariates. In particular, many of the analyses of SC methods explicitly or implicitly refer to units being comparable. This includes the placebo analyses used to test hypotheses \citep{abadie2010synthetic, firpo2018synthetic}. In addition, many applications informally make reference to such assumptions
to justify the inclusion of units
\citep{coffman2012hurricane, cavallo2013catastrophic,liu2015spillovers}.

For observational settings, our insights fall into four categories.
First, we propose  a new estimator (the MUSC estimator) that comes with additional robustness guarantees relative to the previously proposed SC estimators. Second, we develop new  approaches to inference in the form of an unbiased estimator for the finite sample variance. 
Third, the design perspective highlights the importance of the choice of estimand  for inference.
Fourth, we show that the criterion for choosing the weights has some optimality conditions under exchangeability of the treated period  \citep[see also][]{chen2022synthetic}.
We note that our results complement, but do not replace, model-based analyses of the properties of the SC estimator.

In this article, we build on the general SC literature started by
 \citet{abadie2003,abadie2010synthetic, abadie2014}. See \citet{abadie2019using} for a recent survey.
 We specifically contribute to the literature proposing new estimators for this setting, including
  \citet{doudchenko2016balancing}; \citet{abadie2017penalized}; \citet{ferman2017placebo};  \citet{arkhangelsky2019synthetic}; \citet{li2020}; \citet{ben2018augmented}; \citet{athey2017matrix}.
We  also
  contribute to the literature on inference for SC estimators, which includes
 \citet{abadie2010synthetic}; \citet{ doudchenko2016balancing}; \citet{ ferman2017placebo}; \citet{ hahn2016}; \citet{ lei2020conformal}; \citet{ chernozhukov2017exact}.
Furthermore, we add to the general literature on randomization inference for causal effects, \citep[\textit{e.g.}][]{neyman1923, rosenbaum_book,  imbens2015causal, abadie2020sampling, Rambachan2020-ob, roth2021efficient}. In particular, the discussion on the choice of estimands and its implications for randomization inference in \citet{sekhon2020inference} is relevant.
Finally,
we relate to a literature on regression adjustments in randomized experiments \citep{lin}.

\section{SETUP}

We consider a setting with $N$ units, for which we observe outcomes $Y_{it}$  for  $T$ time periods, $i=1,\ldots,N$, $t=1,\ldots,T$. There is a binary treatment denoted by  $W_{it}\in\{0,1\}$, and  a pair of potential outcomes $Y_{it}(0)$ and $Y_{it}(1)$ for all unit/period combinations \citep{rubin1974estimating, imbens2015causal}. 
 The notation assumes there are no dynamic effects, although this would only change the interpretation of the estimand.
There are no restrictions on the time path of the potential outcomes. The $N\times T$ matrices of treatments and potential outcomes are denoted by $\bw$, $\byn$ and $\bye$ respectively. 
The realized/observed outcome matrix is $\by$, with typical element
$\yit\equiv \wit \yite+(1-\wit)\yitn.$
In contrast to most of the SC literature (with \citealp{athey2018design} an exception), we take the potential outcomes  $\byn$ and $\bye$ as fixed in our analysis, and treat the assignment matrix $\bw$ (and thus the realized outcomes $\by$) as stochastic.
For ease of exposition, we focus primarily on the case with a single treated unit and a single treated period.
Many of the insights carry over to the case with a block of treated unit/time-period pairs, see 
Appendix D.1.

In order to separate out the assignment mechanism into the selection of the time period treated  and the unit
treated we write
\[ \bw=\bu\bv^\top,\]
where $\bu$ is an $N$-vector with typical element $U_i\in\{0,1\}$, $\sum_{i=1}^N U_i=1$, and
$\bv$ is a $T$-vector with typical element $V_t\in\{0,1\}$, $\sum_{t=1}^T V_t=1$, satisfying $V_t=\sumi \wit$, $U_i=\sumt \wit$.

\subsection{Estimands}\label{estimands}

 Our primary focus  is on the causal effect for the single treated unit/time-period:
\begin{equation}\label{taut} \tau\equiv \tau(\bu,\bv)\equiv\sumi\sumt U_i V_t\Bigl(\yite-\yitn\Bigr).\end{equation}
For the  case with multiple treated units or periods discussed in Appendix D.1, this estimand can be generalized to the average effect for all the treated unit/time-periods.

There are three other estimands that one might consider. 
First, the average effect for all $N$ units in the treated period, which we call the ``vertical'' effect:
$  \tau^\vvv\equiv\frac{1}{N} \sumi\sumt V_t (\yite-\yitn).$
Second,
the average effect for the treated unit over all $T$ periods, which we call the ``horizontal'' effect:
$ \tauh\equiv\frac{1}{T} \sumi\sumt U_i (\yite-\yitn).$
 Finally, the population average treatment effect:
$ \taup\equiv\frac{1}{NT}\sumi\sumt (\yite-\yitn).$
Which of these estimands is of primary interest depends on the applicaton. If the treatment effect is constant of course the four estimands are all identical, and there is no reason to choose.
In this manuscript we focus on $\tau$, rather than these other average causal effects although the insights obtained for $\tau$ also apply to the other estimands.

If the unit (or time period) treated is  selected completely at random, then $\tau$ itself is unbiased for $\tauv$ (or $\tauh$), and by extension any  estimator that is unbiased for $\tau$ is also unbiased for $\tauv$ (or $\tauh$). However, as an estimator for $\tauv$ (or $\tauh$) it potentially has a different variance  than as an estimator for $\tau$.

\subsection{Assumptions}

In order to derive properties of the estimators, most of the SC literature   uses  a  latent-factor model for the control outcome
\[ Y_{it}(0)=\gamma_{i}'\beta_{t} + \varepsilon_{it} = \sum_{r=1}^R \gamma_{ir}\beta_{tr}+\varepsilon_{it},\]
here with $R$ latent factors
in combination with independence assumptions on the noise components $\varepsilon_{it}$  \citep{abadie2010synthetic, athey2017matrix, amjad2018robust, xu2017generalized}. We focus instead on {design} assumptions, that is, assumptions about the assignment process that governs  the distribution of $\bw$ (or, equivalently, the distributions of $\bu$ and $\bv$) without placing restrictions on the potential outcomes.
Design-based, as opposed to model-based, approaches have a  long tradition in the experimental  literature \citep[\textit{e.g.}][]{fisher1937design, neyman1923, imbens2015causal, rosenbaum_book, cunningham2018causal}, as well as more recently in regression settings \citep{ abadie2020sampling, athey2018design, Rambachan2020-ob}. However, these methods have not yet been used to analyze the properties of SC estimators.

First, we consider random assignment of the units to treatment.
\begin{assumption}\label{ass_random_unit}{\sc (Random Assignment of Units)}
	\[ {\mathbb{P}}(\bu=\buu|\bv=\bvv)=
	\left\{
	\begin{array}{ll}\frac{1}{N}\hskip1cm & \mathrm{if}\ \ u_i\in\{0,1\}\: \forall i,\ \  \sumi u_i=1,\\
		0 & \mathrm{otherwise.}
	\end{array}\right.\hskip0.3cm {\rm or}\ 
	 \ \bu\ \indep\ (\by(0),\by(1),\bv).\]
\end{assumption}

Because SC methods are typically used in observational settings, this assumption may seem unusual to invoke for an SC setting. However, many SC applications implicitly use unit randomization assumptions when implementing placebo tests {\it e.g.}, \citet{abadie2010synthetic}. 
Moreover,  random treatment assignment after adjusting for observed covariates is an assumption underlying many causal analyses. Studying SC methods under randomization can also improve estimation and inference for true randomized settings, especially in cases where the number of treated units is small as is typical when using SC.

Second, we consider the assumption  that the  treated period was randomly selected from the $T-N-1$ periods under observation after the first $N+1$ observations. (We do not allow the treatment to occur during the first $N+1$ periods to avoid having insufficient data to calculate the Synthetic Control weights without regularization.) 
\begin{assumption}\label{ass_random_time}{\sc (Random Assignment of Treated Period)}
	\[ {\mathbb{P}}(\bv=\bvv|\bu=\buu)=
	\left\{
	\begin{array}{ll}\frac{1}{T-N-1}\hskip1cm & \mathrm{if}\ \ v_t\in\{0,1\} \: \forall t\geq N+2,\ \ \sum_{t=N+2}^T v_t=1,\\
		0 & \mathrm{otherwise.}
	\end{array}\right. \hskip0.3cm {\rm or}\ 
	 \ \bv\ \indep\ (\by(0),\by(1),\bu).\]
\end{assumption}
Although this assumption is not plausible in many cases, as it is often only the last period(s) that are treated, it is useful to consider its implications. It formalizes the often implicit SC assumption that there is a
within-period relationship between control outcomes for different units that is stable over time. See also the discussion in \citet{chen2022synthetic}.

Most of our discussion concerns finite-sample results, imposing mainly \autoref{ass_random_unit} (random treated unit). However, for some results it will be useful to consider large-$T$ approximations. 
For large-$T$ results, 
first 
 define
$\by_{\cdot t}(0)$ to be the $N$ vector with typical element $Y_{it}(0)$. Define
the averages up to period $T$ of the first and the centered second moment:
\[ \hat{\mathbf{\mu}}_T\equiv \frac{1}{T}\sum_{t=1}^T \by_{\cdot t}(0),
\hskip1cm \hat{\mathbf{\Sigma}}_T\equiv\frac{1}{T}\sum_{t=1}^T
\left(\by_{\cdot t}(0)- \hat{\mathbf{\mu}}_T\right)
\left(\by_{\cdot t}(0)- \hat{\mathbf{\mu}}_T\right)^\top.
\]
\begin{assumption}\label{stationarity}{\sc (Large-$T$ Stationarity)}
For some finite $\mathbf{\mu}$ and finite positive-definite $\mathbf{\Sigma}$, the sequence of $\by_{\cdot t}(0)$ satisfies, as $T \rightarrow \infty$,
\[ \hat{\mathbf{\mu}}_T\longrightarrow \mathbf{\mu},\hskip1cm 
\hat{\mathbf{\Sigma}}_T\longrightarrow \mathbf{\Sigma}.\]
\end{assumption}

\subsection{Generalized Synthetic Control Estimators}

In this section, we introduce a class of SC-type estimators. 
This class, which we refer to as Generalized Synthetic Control (GSC) estimators, includes the DiM estimator, the original SC estimator proposed by \citet{abadie2010synthetic}, and three modifications as special cases.
For the purpose of a randomization-based analysis, we must define these
 estimators for all possible treatment assignment vectors $\bu$ and $\bv$, not just the realized assignment. 
 
 \subsubsection{Estimators}
 
 We characterize the GSC  estimators in terms of a set of weights $M_{ijt}$,
 indexed by $i=1,\ldots,N$, $j=0,\ldots,N$, and $t=1,\ldots,T$.
  Given a set of  weights $\momega$, treatment assignments $\bu$, $\bv$, and outcomes $\by$,
 the GSC estimator has the form
 \begin{equation}\label{general_estimator} \hat\tau^{\rm GSC}=\hat\tau(\bu,\bv,\by,\momega)\equiv
 \sumituv\left\{M_{i0t}+ \sum_{j=1}^N M_{ijt} Y_{jt}\right\}.\end{equation}
We show that this estimator is stochastic only through the $U_i$ and $V_t$ by showing below that the weights are non-stochastic.
 
 The estimators in the GSC class differ in the choice of the weights $\momega.$ There are generally two components to this choice.
First, there is an objective function that defines the  weight within the set of possible weights. This objective function is identical for all GSC estimators we consider in the current article.
 Second, there is a non-stochastic set of possible weights, denoted by $\calm$, over which we search for an optimal weight. 
 These sets $\calm$ differ between the estimators we consider, and in fact it is the only way in which the estimators differ. 
 A summary of the differences between the estimators is given in Table \ref{table_gsc_comparisons}.

\begin{table}[!htbp] \centering 
  \caption{GSC Estimator Comparison} 
  \label{table_gsc_comparisons} 
\begin{tabular}{l cccccc}
    \toprule
  & {DiM}    & {SC} & {MSC} & {USC} &   {MUSC}   & {DiD}  \\ \midrule
Uniform Weights & \textrm{Yes} & \textrm{No} & \textrm{No}& \textrm{No}& \textrm{No}&  \textrm{Yes}  \\
Intercept &\textrm{No}  &\textrm{No}  &  \textrm{Yes}  &\textrm{No} &  \textrm{Yes}  &  \textrm{Yes}  \\
Column Weights Sum to Zero &\textrm{Yes}  &\textrm{No}  & \textrm{No}  &  \textrm{Yes}  &  \textrm{Yes}  &  \textrm{Yes}  \\
\bottomrule
\end{tabular}\\ \vskip0.2cm
{\footnotesize DiM: Difference in Means estimator; SC: Synthetic Control estimator of
\cite{abadie2010synthetic}; \\ MSC: Modified Synthetic Control estimator of \cite{doudchenko2016balancing}; USC: Unbiased Synthetic Control; MUSC: Modified Unbiased Synthetic Control estimator; DiD: Difference-in-Differences estimator.}
\end{table}

All sets of weights for the different estimators are subsets of the following set:
\begin{equation}\label{setweights} \calm^0=\left\{\momega\left|
M_{iit}=1,\forall i\geq 1, \forall t;
 M_{ijt}\leq 0, \forall i,j\geq 1, i \neq j,\forall t;
 \sum_{j=1}^{N}M_{ijt}=0\: \forall i\geq 1, \forall t
 \right.
\right\}
.\end{equation}
There are three restrictions captured in this set. First,  the weight for the treated unit is equal to one.
Second, the weight for unit $j$ for the prediction of the causal effect for unit $i$ is nonpositive:
\begin{equation}\label{nonpositive} M_{ijt}\leq 0,\ \: \forall i\in \{1,\ldots,N\},j\in \{1,\ldots,N\} \setminus \{i\},t\in \{1,\ldots,T\}.\end{equation}
The third restriction requires that the weights for all units in the prediction for the causal effect for unit $i$ in period $t$ sum to zero. Because the weight for unit $i$ in this prediction is restricted to be equal to one, this means that the weights for the control units sum to minus one.
We consider four estimators in this class, characterized by four sets of possible weights $\calm\subset\calm_0$, described in Section 2.3.3.

 \subsubsection{The Objective Function}
 
 We start with the second component of the choice of weights, the objective function.
 For a given matrix of outcomes $\by$, and a given set of possible weights $\calm$,  define the tensor $\momega(\byn;\calm)$ with elements $M_{ijt}$, as
 \begin{equation}\label{eq:gen_sc}\momega(\byn;\calm)\equiv\argmin_{\momega\in\calm} \sumit\left\{\sum_{s< t} \left(M_{i0t}+\sumj M_{ijt} Y_{js}(0)\right)^2\right\}. \end{equation}
 As long as the sets $\calm$ are non-stochastic, the definition of the weights implies they are non-stochastic, and so that the estimators we consider are stochastic only through the assignment vectors $\bu$ and $\bv$. We only sum over $s<t$ to ensure that the weights only depend on pre-treatment periods.

What is the motivation for the objective function in (\ref{eq:gen_sc})?
The expected squared error of the estimator $\hat\tau^{\gsc}$, under unit and time randomization (Assumptions \ref{ass_random_unit} and \ref{ass_random_time}),  is \[\frac{1}{N(T-N-1)}\sum_{i=1}^N \sum_{t=N+2}^T \left(M_{i0t}+ \sum_{j=1}^N M_{ijt} Y_{jt}(0)\right)^2.\]
We cannot evaluate this squared loss because it depends on $Y_{it}(0)$ that we do not observe. However, we can 
use the analogue from {the control periods} before the treated period.
With sufficiently large $T$, these  values are comparable to the current value, suggesting the objective function (\ref{eq:gen_sc}).
 
\subsubsection{Feasible Weights}

 The DiM estimator
  corresponds to the  case with
  \[ \calm^{\rm DiM}=\biggl\{\momega\in\calm^0\biggl|
M_{i0t}=0
 \: \forall i,t, M_{ijt}=-1/(N-1)\: \forall i\neq j,t\biggr\}.\]
 Relative to DiM, the DiD estimator relaxes the no-intercept restriction,
   \[ \calm^{\rm DiD}=\biggl\{\momega\in\calm^0\biggl|M_{ijt}=-1/(N-1)\: \forall i\neq j,t\biggr\}.\]
The original  SC estimator \citep{abadie2003, abadie2010synthetic} corresponds to the estimator based on (\ref{eq:gen_sc}) with the set $\calm$ defined as the subset of $\calm^0$ satisfying
\[ \calm^\ssc=\biggl\{\momega\in\calm^0\biggl|
M_{i0t}=0
 \: \forall i,t\biggr\}.\]
 
The modification introduced in \citet{doudchenko2016balancing} and \citet{ferman2019synthetic} allows for an intercept by dropping the  restriction $M_{i0t}=0$, leading to the Modified Synthetic Control (MSC) estimator with 
$ \calm^\msc=\calm^0.$ \citet{arkhangelsky2019synthetic}
show that the inclusion of the intercept can be interpreted as including a unit fixed effect in the regression function. In \autoref{properties} we discuss how the inclusion of the intercept ties in with the time randomization assumption. The presence of the intercept also reduces the importance of time-invariant covariates.

A second modification of the basic SC estimator, the Unbiased Synthetic Control (USC) estimator,  places an additional set of restrictions on the weights beyond those used in the SC estimator, namely that  all units are in expectation used as controls as often as they are used as treated units:
\[ \calm^\usc=\biggl\{\momega\in\calm^\ssc\biggl|\sum_{i=1}^NM_{ijt}=0\: \forall t, \forall j\geq 1\biggr\}.\]
Finally, we combine the two modifications of the SC estimator, the relaxation of the constraint that the  intercept is zero and the  additional restriction on the adding up of the control weights, leading to our main alternative to the SC estimator, the Modified Unbiased Synthetic Control (MUSC) estimator:
\begin{equation}\label{eq:musc} \calm^\musc=\biggl\{\momega\in\calm^0\biggl| \ \sum_{i=1}^NM_{ijt}=0\: \forall t,\forall j \geq 1\biggr\}.\end{equation}

These four sets of restrictions define four estimators.
Our focus is primarily on $\hat\tau^\ssc$
and $\hat\tau^\musc$, while comparisons with the intermediate cases $\hat\tau^\msc$ and $\hat\tau^\usc$ serve to aid the interpretation of the two restrictions that make up the difference between  $\hat\tau^\ssc$
and $\hat\tau^\musc$.

As a comparison, we also consider an additional modification of the SC estimator that relaxes the assumption that the weights for the controls sum to one, leading to the SC--NR (Synthetic Control -- no restriction) estimator:
\[\calm^{\textnormal{SC--NR}}=\left\{\momega\left|
M_{iit}=1,\forall i\geq 1, \forall t
 \right.
\right\}
.\]

\section{PROPERTIES}\label{properties}

In this section, we investigate the formal properties of the various estimators given the unit and/or time randomization assumptions. 
\autoref{motivating example} provides a preview of our main result on the bias of the standard Synthetic Control estimator under unit randomization.
\autoref{sec: SC bias} characterizes the bias of GSC estimators and gives a condition for unbiasedness. \autoref{sec: variance} calculates the variance for GSC estimators and gives an unbiased estimator for the variance, which \autoref{sec: placebo variance} compares to the common placebo variance estimator; \autoref{sec: DiM improvement} compares the design-based GSC variance to other estimators. Finally, \autoref{sec: network} gives a network interpretation of GSC estimators, which \autoref{sec: propensity} relates to non-constant propensity scores.

\subsection{A Motivating Example} \label{motivating example}

\begin{figure}
    \colorlet{unit1}{red!50!black}
    \colorlet{unit2}{green!50!black}
    \colorlet{unit3}{blue!50!black}

    \newcommand{\makecoordinates}{
    \useasboundingbox (1,1) rectangle (8,-1.2);
            \draw[->] (1,0) -- (7,0) node[right]{$Y_{i1}$};
            \draw[color=unit1] (1 * 2,.2) -- (1 * 2,-.2) node[below]{AZ} ;
                \coordinate (1) at (1 * 2,0);
            \draw[color=unit2] (2 * 2,.2) -- (2 * 2,-.2) node[below]{CA} ;
                \coordinate (2) at (2 * 2,0);
                \draw[color=unit3] (3 * 2,.2) -- (3 * 2,-.2) node[below]{NY} ;
                \coordinate (3) at (3 * 2,0);
            \coordinate (offset) at (0,0.2);
            \coordinate (0) at (1.2,0);

    }

    \centering
    
    \begin{subfigure}[b]{0.3\textwidth}
        \centering

        \scalebox{0.70}{
        \begin{tikzpicture}
            \makecoordinates{}
        \end{tikzpicture}
        }
        \caption{Pre-treatment outcomes}
    \end{subfigure}
    \begin{subfigure}[b]{0.3\textwidth}
        \centering

              \scalebox{0.70}{\begin{tikzpicture}
            \makecoordinates{}
           
            \draw[-{Latex[scale=2]},color=unit2] (2) to[bend right] node[above]{\scriptsize 1} (1);
            \draw[-{Latex[scale=2]},color=unit2] (2) to[bend right] node[below]{\scriptsize 1} (3);
            
            \draw[-{Latex[scale=2]},color=unit1] (1) to[bend right] node[below]{\scriptsize \sfrac{1}{2}} (2);
            \draw[-{Latex[scale=2]},color=unit3] (3) to[bend right] node[above]{\scriptsize \sfrac{1}{2}} (2);

        \end{tikzpicture}}
        \caption{Weights, SC estimator}
    \end{subfigure}     
    \begin{subfigure}[b]{0.3\textwidth}
            \centering
              \scalebox{0.70}{\begin{tikzpicture}
            \makecoordinates{}
            
            \draw[-{Latex[scale=2]},color=unit1] (1) to[bend right] node[below]{\scriptsize \sfrac{1}{2}} (2);
            \draw[-{Latex[scale=2]},color=unit3] (3) to[bend right] node[above]{\scriptsize \sfrac{1}{2}} (2);
            
            \draw[-{Latex[scale=2]},color=unit2] (2) to[bend right] node[above]{\scriptsize \sfrac{1}{2}} (1);
            \draw[-{Latex[scale=2]},color=unit3] (3) to[bend right=55] node[above]{\scriptsize \sfrac{1}{2}} (1);
            
            \draw[-{Latex[scale=2]},color=unit2] (2) to[bend right] node[below]{\scriptsize \sfrac{1}{2}} (3);
            \draw[-{Latex[scale=2]},color=unit1] (1) to[bend right=55] node[below]{\scriptsize \sfrac{1}{2}} (3);
            
            \coordinate (FIRST NE) at (current bounding box.north east);
            \coordinate (FIRST SW) at (current bounding box.south west);
            
        \end{tikzpicture}}
        \caption{Weights, USC estimator}
    \end{subfigure}

    \caption{Pre-treatment outcome in three-unit, two-period example. An outgoing arrow represents the weight assigned to that unit when the target of the arrow is treated, and arrows are colored by the unit the respective weight is put on.}
    \label{fig:ABC2}
\end{figure}
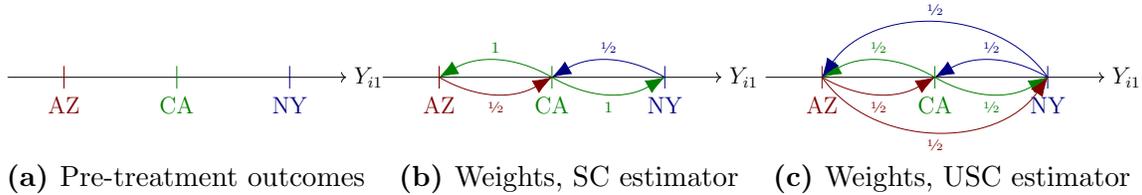

In this section we
 preview the result that the Synthetic Control  estimator is biased and that the bias can be removed by restricting the SC weights in a simple example with three units (say, Arizona, AZ; California, CA; New York, NY) and two periods ($t \in \{1,2\}$), where treatment is assigned to a single unit in the second time period with equal probability for each unit. 
  The three pre-treatment  outcomes are depicted in Panel (a) of Figure \ref{fig:ABC2}.  In this setting we compare the Synthetic Control (SC) estimator and an  unbiased version of the SC estimator  (USC). 
 
  If CA is treated, the SC estimator puts equal weight on each of the equidistant control units. When either of the peripheral units, AZ or NY, is treated, the SC estimator puts all its weight on CA. The weights of the standard SC estimator are represented in Panel (b). The result is that in expectation CA is used as a control unit more than AZ or NY, and the total weight for CA as a control  over   all three assignments exceeds the weight CA gets as a treated unit. 
This difference in weights, or equivalently the imbalance of CA's use as treatment and control unit, is what creates bias in the SC estimator under randomization. Adding a simple constraint to the weights which rules out this imbalance yields an unbiased Synthetic Control  estimator. The resulting weights can be found in Panel (c). 
In this three-unit example, the  unbiased SC estimator is simply the difference-in-means estimator, but with more than three units this is not generally true. 

\subsection{The Bias of the GSC Estimators}\label{sec: SC bias}

Having given a simple example of the bias of the SC estimator, we now study more broadly the bias of the four GSC estimators relative to the treatment effect for the treated unit,
$\tau$. 
We summarize the results in \autoref{tab:bias_sc}.

\begin{table}[ht]
	\caption{Bias Properties of GSC Estimators for $\tau$}
	\centering
	\small
	\begin{tabular}{lccccc}
		\toprule
		 & \multicolumn{5}{c}{\textbf{Maintained Assumptions}}\\ 
		 \cmidrule(lr){2-6}
	 Randomization & Unit & Time & Time & Unit \& Time & Unit \& Time \\
		&  & & Large $T$
	& & 
	Large $T$\\ 
	 Assumptions 	& \ref{ass_random_unit}
	& \ref{ass_random_time}
	& \ref{ass_random_time}, \ref{stationarity}
	& \ref{ass_random_unit}, \ref{ass_random_time}
	& \ref{ass_random_unit}, \ref{ass_random_time}, \ref{stationarity}
	\\
    \midrule
	SC & \bias & \bias & \bias & \bias & \bias\\	
	MSC & \bias & \bias &\unbias & \bias& \unbias\\	
	USC & \unbias & \bias &\bias & \unbias& \unbias\\	
MUSC & \unbias & \bias & \unbias& \unbias &  \unbias \\
		\bottomrule
	\end{tabular}

	\label{tab:bias_sc}
\end{table}

Recall the general definition of the GSC  estimators  in (\ref{general_estimator}), which can be rewritten as
\[\hat\tau(\bu,\bv,\by,\momega)
=\sumt  \sumi V_t U_i\left\{\yite+M_{i0t}+ \sum_{j=1}^N (1-U_j) M_{ijt} Y_{jt}(0)\right\}.
\]
The estimation error relative to the treatment effect for the treated is equal to
\[\hat\tau(\bu,\bv,\by,\momega)-\tau(\bu,\bv)=
\sumituv\left\{\yitn+M_{i0t}+\sum_{j=1}^N (1-U_j) M_{ijt} Y_{jt}(0)\right\}.
\]
We consider the bias of the GSC estimators separately for the estimators without an intercept (the SC and USC estimators), and for the estimators with the intercept 
(\ref{eq:gen_sc}) (the MSC and MUSC estimators).
\begin{lemma}
	Suppose \autoref{ass_random_unit} (random assignment of units to treatment) holds. If one of the following two conditions holds:\\
	$(i)$ the intercept is zero, $M_{i0t}=0$ for all $i,t$, \\
	or $(ii)$ the intercept is not constrained and estimated through (\ref{eq:gen_sc}),\\
	then the conditional (on $\bv$) bias vanishes if the set of weights  $\calm$ imposes the condition that
\begin{equation}\sumi  M_{ijt}=0\ \: \forall j=1,\ldots,N,t=1,\ldots,T.
\label{addingup}
\end{equation}
\end{lemma}

This lemma shows that $\tau^\usc$ and $\tau^\musc$ are unbiased, because both estimators only search over weight sets that satisfy the adding-up condition in (\ref{addingup}).
The above formulas also immediately lead to the bias of the SC estimator under \autoref{ass_random_unit}.
\begin{lemma}
	Suppose \autoref{ass_random_unit} (random assignment of units to treatment) holds. Then conditional on $\bv$, the bias of the SC estimator is 
\begin{equation} \label{bias} \mathrm{Bias}^\ssc=\mme\left[\left.\hat\tau-\tau\right|\bv\right]=
	\frac{1}{N} \sum_{j=1}^N \sum_{t=1}^T V_t Y_{jt}(0) \sum_{i=1}^N M^\ssc_{ijt} .
\end{equation}
\end{lemma}
The intuition for the bias of the SC estimator also holds for the simple matching estimator \citep[{\it e.g.}, ][]{abadie2006}, which is generally biased under randomization in finite samples.

In principle one can
 estimate the bias  for the SC estimator in equation (\ref{bias}) and generate an unbiased estimator by subtracting the estimated bias from the standard SC estimator. However, in simulations, the bias estimates are very imprecise and as a result the properties of this de-biased estimator are not  attractive in terms of RMSE.

To see the role that time randomization and the presence of an intercept play in the bias, consider the MSC estimator in a setting with large $T$ and random selection of the treated period. For ease of exposition, suppose unit $N$ is the treated unit.
One can view the MSC estimator as a regression estimator where we regress the outcomes $Y_{N1},\ldots,Y_{NT}$ on the treatment indicator, the predictors $Y_{1,t},\ldots,Y_{N-1,t}$, and an intercept.
It is well known that this leads to an estimator that is asymptotically unbiased in large samples \citep[in this case meaning large $T$; see][]{freedman2008regression,lin, imbens2015causal}. 

A final comment concerns the magnitude of the bias. Although in our illustration the bias is small, it can in fact be arbitrarily large. Consider a case with binary outcomes, $Y_{it}(w)\in\{0,1\}$ and the treatment effect equal to zero for all units, $Y_{it}(0)=Y_{it}(1)$. Suppose the number of time periods is equal to the number of units. Moreover, suppose that for the first unit $Y_{1t}=0$ for $t=1,\ldots,T$. For all  other units $i\neq 1$ $Y_{it}(0)=Y_{it}(1)=0$ for $t\notin\{i,T\}$ with $Y_{ii}(0)=Y_{ii}(1)=Y_{iT}(0)=Y_{iT}(1)=1$. In that case the first unit is matched with equal weight to all other units, $M_{ijT}=-1/(N-1)$ for j=$2,\ldots,N-1$, and all other units are matched to the first unit: $M_{i1T}=-1$ for $j=2,\ldots,N$. The bias in this case is $(N-2)/N$ which can be made arbitrarily close to the maximum possible value of 1.

\subsection{The Exact GSC Variance and its Unbiased Estimation}\label{sec: variance}

Here we analyze the GSC estimator under unit randomization (\autoref{ass_random_unit}) only, conditioning on the time treated. Alternatively, we could analyze the GSC estimator under time randomization only, or under both unit and time randomization, but those inferences may be less attractive in practice if only the last period is the treated period.
\begin{lemma}
 Suppose \autoref{ass_random_unit} holds. Then
 \[\mmv(\bv,\momega)=\mme
 \left[\left.\left(\hat\tau(\bu,\bv,\by,\momega)-\tau\right)^2\right|\bv\right] =\frac{1}{N}\sumit V_t \left(
 M_{i0t}+\sum_{j=1}^N M_{ijt} Y_{jt}(0)
 \right)^2.\]
\end{lemma}
For the unbiased estimators, this is the variance around the treatment effect $\tau$ on the treated unit, while for the other estimators, it is the expected squared error.
The challenge is that the variance depends on control outcomes that we do not observe. 
However, we can estimate this variance without bias under unit randomization.

\begin{prop}\label{prop1} Suppose \autoref{ass_random_unit} holds. Then the estimator
    \begin{equation}\label{var}
        \hmmv
        =\sumituv\left\{\frac{1}{N{-}3}
         \sum_{\substack{k = 1 \\ k \neq i}}^N \left( \sum_{\substack{j = 1 \\ j \neq i}}^N M_{kjt} (Y_{kt} {-} Y_{jt})\right)^2 
         - \frac{1}{(N{-}2)(N{-}3)} \sum_{\substack{k = 1 \\ k \neq i}}^N
         \sum_{\substack{j = 1 \\ j \neq i}}^N
         \momega^2_{kjt} (Y_{kt} {-} Y_{jt})^2
         \right.
         \end {equation}
         \[
      \left. \hskip3cm +
        \frac{2}{N{-}2} \sum_{\substack{k = 1 \\ k \neq i}}^N M_{k0t} \left( \sum_{\substack{j = 1 \\ j \neq i}}^N M_{kjt} (Y_{jt} {-} Y_{kt} )\right)
        +
        \frac{1}{N} \sum_{k = 1}^N \momega^2_{k0t}
         \right\},
    \]
    is unbiased for $\mmv(\bv,\momega)$.
\end{prop}
This result may be somewhat surprising. Note that in completely randomized experiments there is no unbiased estimator of the variance of the simple difference-in-means estimator for the average treatment effect \citep{imbens2015causal}. The current result is different because here we focus on the effect for the treated only. For that estimand, there is an unbiased estimator for the variance of the difference-in-means estimator in the case of randomized experiments \citep[{\it e.g.},][]{sekhon2020inference}.

The variance estimator in this proposition has three terms.
The first takes the form of a leave-one-out estimator based on the control units excluding the treated unit. The remaining two terms
 correct for over-counting the diagonal elements in the inner square of the first term and additional terms for the intercept.
In the special case of the DiM estimator, the variance reduces to the standard variance estimator. In that case, it is guaranteed to be non-negative, which does not hold in general.

\subsection{The Placebo Variance Estimator} \label{sec: placebo variance}

To put the proposed variance estimator in (\ref{var}) in perspective, 
 we consider here an alternative approach for estimating the variance of the SC estimator. Versions of this placebo variance estimator have  been proposed previously both for testing zero effects \citep[\textit{e.g.}][]{abadie2010synthetic} and for constructing confidence intervals \citep[\textit{e.g.}][]{doudchenko2016balancing}.
Suppose unit $i$ is the treated unit. 
We put this unit aside, and focus on the $N-1$ control units.
For each of these $N-1$ control units (indexed by $j=1,\ldots,N, j\neq i$), we recalculate the weights, leaving out the treated unit, and then estimate the treatment effect.
For ease of exposition we focus on the case where the last period is the treated period, $V_T=1$.

We now define $(N-1)\times N$ weight matrices and sets of weight matrices $\momegai$ and $\calmi$ from the restriction of $\momega$ and $\calm$ to units $j \neq i$.
The weights are defined as
\begin{equation}\label{gen_pcb}\momegai(\by,\calmi)=\argmin_{\momegai\in\calmi} \sum_{s=1 }^{T-1}
\sum_{\substack{j=1 \\ j \neq i}}^N
\left(M^{(i)}_{j0}+\sum_{\substack{k=1 \\ k \neq i}}^N M^{(i)}_{jk} Y_{ks}\right)^2. \end{equation}
Given these weights,
the placebo estimator is 
\[ \hat\tau_j^{(i)}=M^{(i)}_{j0}+\sum_{\substack{k=1 \\ k \neq i}}^N M^{(i)}_{jk} Y_{kT},\hskip1cm {\rm for}\ j\neq i.\]
Because unit $j$ is a control unit, this is an estimator of zero, and 
the placebo variance estimator uses it to estimate the variance of $\hat\tau$ as
\[ \hat\mmv^{\pcb}=\frac{1}{N-1}\sumi U_i  \sum_{\substack{j=1 \\ j \neq i}}^N 
\left(\hat\tau_j^{(i)}\right)^2\
=\frac{1}{N-1}\sumi U_i \sum_{\substack{j=1 \\ j \neq i}}^N 
\left(
 M^{(i)}_{j0}+\sum_{\substack{k=1 \\ k \neq i}}^N M^{(i)}_{jk} Y_{kT}
 \right)^2.\]
 This variance estimator can be upward as well as downward biased, depending on the potential outcomes. In order to demonstrate this, we provide two toy examples in Appendix B where the placebo variance estimator is biased downward and upward, respectively.

\subsection{Randomization Inference and Confidence Intervals}

\label{subsec:randomization}

Within our design-based framework, a natural way of testing and providing confidence intervals is based on performing randomization inference directly, rather than relying solely on the estimated variance.
In this section, we lay out how one can construct randomization-based tests and confidence intervals, building upon placebo tests for Synthetic Control in \cite{abadie2010synthetic} and similar to \cite{firpo2018synthetic}.
As in our related discussion of the placebo variance in the previous section, we focus on the case of unit randomization (\autoref{ass_random_unit}) with treatment in the last period ($V_T = 1$).
We note that the derivation applies to any GSC estimator, not just the specific USC and MUSC estimators.

We consider tests of the null hypothesis
$\tau_{i} = Y_{iT}(1) - Y_{iT}(0)= \beta$, where $i$ is the index of the treated unit.
In our setting, for every unit $j$,
$\hat{\tau}_j-\tau_j = M_{j0T} + \sum_{k=1}^N M_{j k T} Y_{kT}(0)$ where $\hat{\tau}_j$ is the GSC estimator of $\tau_j = Y_{jT}(1) - Y_{jT}(0)$.
Under the null hypothesis, 
\(Y_{jT}(0) = Y_{jT} - \beta \ \mathbbm{1}\{j = i\}.\)
That means that under the null
\[\hat{\tau}_j - \tau_j = M_{j0T} + \sum_{k=1}^N M_{jkT} Y_{kT}-\beta \ M_{jiT}.\]

We consider a test based on quantiles of $\hat{\tau}_j-\tau_j$.
We have for $j \neq i$ that
\begin{align*}
    &
    &
    &\hat{\tau}_j-\tau_j \geq \hat{\tau}_{i}-\tau_{i}
    \\
    &\iff
    &
    &\beta \: (M_{i i T}-M_{j i T})+
    (M_{i0T}-M_{j0 T})
    + \sum_{k=1}^N (M_{jkT}-M_{ik T}) Y_{kT} \geq 0\\
    &\iff
    &
    &\beta  \geq \frac{
    M_{i0T}-M_{j0 T} +
    \sumj (M_{i k T} - M_{jkT}) Y_{kT}}{M_{i i T}-M_{j i T}} = \hat{\beta}_j.
\end{align*}
We specifically construct a permutation test of size $\alpha$ for $\hat{\tau}_j-\tau_j$.
With a two-tailed test of size $\alpha$, we would not reject $H_0: Y_{iT}(1) = Y_{iT}(0) + \beta$ whenever $q_{\alpha/2}(\sum_{j=1}^N U_j (\hat{\tau}_j-\tau_j)) < \hat{\tau}_{i}-\tau_{i} \leq q_{1-\alpha/2}(\sum_{j=1}^N U_j (\hat{\tau}_j-\tau_j))$, which is equivalent to
\[
    q_{\alpha/2}
    \left(\sum_{j=1}^N U_j \hat{\beta}_j \right) < \beta \leq q_{1-\alpha/2}\left(\sum_{j=1}^N U_j \hat{\beta}_j\right),
\]
where we set $\hat{\beta}_i = \beta$ and choose randomized quantiles to ensure exact size.
This procedure yields a randomization test of $H_0$ that has exact size within our design-based framework.
It differs from the test considered in \cite{firpo2018synthetic}, which is instead based on the fit of the Synthetic Control estimator and uses a weighted $p$-value to test null hypotheses about treatment effects.

We can then obtain confidence intervals based on test inversion, analogous to \cite{firpo2018synthetic} but based on our specific unit permutation test of size $\alpha$ above.
in order to obtain a confidence interval at level $1-\alpha$ for the treatment effect $\tau_i$ on unit $i$.
Writing $\hat{\beta}_{(1)},\ldots, \hat{\beta}_{(N-1)}$ for the order statistics of $\hat{\beta}_j$ with $j \neq i$,
we obtain a $1-\alpha$ confidence interval
\[
    \tau_{i} \in
    \left[
        \hat{\beta}_{(N \alpha /2)},
        \hat{\beta}_{(N (1-\alpha /2))}
    \right]
\]
for the estimand $\tau_i$, which is itself random.
Here, we assume that a fractional order statistic $\hat{\beta}_{(u)}$ with $u = \lfloor u \rfloor + \delta$ for $\delta \in [0,1)$ is equal to $\hat{\beta}_{(\lfloor u \rfloor)}$ with probability $1-\delta$, and $\hat{\beta}_{(\lfloor u \rfloor + 1)}$ with probability $\delta$.
We could alternatively obtain confidence intervals that have potentially shorter length e.g. by inverting a test based on the quantiles of $|\hat{\tau}_j - \tau_j|$.

\subsection{Improvement over the Difference-in-Means Estimator}\label{sec: DiM improvement}

Simulations in \autoref{section:simulations} show that the variance of the MUSC and SC estimators can be substantially smaller than that of the DiM estimator. However, that is not guaranteed if we only make the assumption that the treated unit was randomly selected. It is possible that in the treated period the pattern between the outcomes is very different from that in the other periods, so that the MUSC and SC estimators have variances larger than that of the DiM estimator. However, this scenario can be ruled out if the treated time period is randomly selected among all periods (\autoref{ass_random_time}) and the number of time periods is large (\autoref{stationarity}):

\begin{prop}\label{prop:dominance} Suppose Assumptions \ref{ass_random_unit}--\ref{stationarity} hold. Let $\mmv({\cal M}) =\mme
 \left[\left(\hat\tau-\tau\right)^2\right]$ be the expected squared error of the GSC estimator $\hat{\tau}$ with a time-invariant and convex constraint set ${\cal M}$, and let $\mmv^{\rm DiM}$ be the variance of the corresponding DiM estimator. 
Suppose that the constraint set ${\cal M}$ allows for equal weights that sum to one (the DiM estimator).
Then 
\[\lim_{T\rightarrow \infty} \mmv({\cal M}) \leq \lim_{T\rightarrow \infty} \mmv^{\rm DiM}. \]
\end{prop}

An informal proof goes as follows.
Writing ${\cal B} \subseteq \mathbb{R}^{N \times (N+1)}$ for the constraint set at a given time (so that ${\cal M} = {\cal B}^T$),
define
\[
    \mathbf{\beta}^*_T = \argmin_{\mathbf{\beta} \in {\cal B}}
    \mme\left[\sum_{i=1}^N \sum_{t=1}^T U_i V_t \left(Y_{it}(0)-\beta_{i0}-\sum_{j \neq i}\beta_{ij} Y_{jt}(0)\right)^2 \right]
\]
for the best set of GSC weights from $\cal M$ that are constant over time.
First, since $\cal M$ contains the DiM estimator, $\mathbf{\beta}^*_T$ has expected loss at most that of the DiM estimator.
Second, the weights
of the GSC estimator $\hat{\tau} = \sum_{i=1}^N \sum_{t=1}^T U_i V_t (\widehat{M}_{i0t}+\sum_{j\neq i}\widehat{M}_{ijt} Y_{it})$ 
for large $T$ and sufficiently large $t$ approximate the oracle GSC weights $\mathbf{\beta}^*_T$, and achieve similar loss in the limit.
Note that this  holds for the set of weights ${\cal M}^{\rm SC}$, as well as for other GSC estimators. \citet{chen2022synthetic} generalizes this result and discusses the connection to online learning.

\subsection{A Network Interpretation of SC Estimators and Their Bias} \label{sec: network}

To understand the bias of SC estimators, we note that SC weights
\begin{align*}
\momega \in \calm^{\ssc}=\left\{\momega\left|
M_{iit}{=}1,\forall i,t;
 M_{ijt}{\leq} 0, \forall i, j{\geq} 1,t;
 \sum_{j=1}^{N}M_{ijt}=0\: \forall i,t;
 M_{i0t} {=} 0 \forall i
 \right.
\right\}
\end{align*}
for a given treatment time $t$ can be understood as a directed network with vertices $i$ and edge flows (or weights) $W_{ij} = -M_{jit} \geq 0$ from vertex $i$ to vertex $j \neq i$.
The weight constraint $\sum_{j=1}^{N}M_{ijt}=0$ then ensures that the total incoming flow equals one for all vertices $i$, $\sum_{j \neq i} W_{ij} = 1$.
An example of such a network representation of an SC estimator is given in \autoref{fig:ABC2}(b).

Bias arises in the SC network whenever the incoming flow (which measures how often a unit is treated) is not the same as the outgoing flow of a vertex (which measures how often each unit is used as a control). The network corresponding to the SC estimator in \autoref{fig:ABC2} is imbalanced: for the outside vertices, inflow exceeds outflow, while the inside vertex has higher outflow than inflow.
Imposing the unbiasedness constraint $\sum_{j=1}^NM_{jit}=0$ is equivalent to imposing the flow balance constraint
$
    \sum_{j \neq i} W_{ij} = \sum_{j \neq i} W_{ji}
$
at all vertices $i$, ensuring that the corresponding units are used as often as controls as they are treated.
Such a network is obtained in \autoref{fig:ABC2}(c), where inflows and outflows are balanced.

Beyond providing an intuitive language to represent Synthetic Control estimators, we show in \autoref{sec: propensity} how tools from network analysis can help analyzing their properties.
There, we show that the eigenvector centrality in the network represented by $\mathbf{W}$ relates to propensity scores subject to which an SC estimator is unbiased.
With this network representation, we thus connect the SC estimator to the tools and insights from the literature on networks across statistics and the social sciences \citep[\textit{e.g.},][]{Jackson2010-tw, De_Paula2020-ux}.

\subsection{Non-Constant Propensity Scores}
\label{sec: propensity}

So far we have assumed that treatment is assigned with equal probability across units, time periods, or unit--time pairs.
Yet the theory we develop generalizes to non-constant propensity scores.
See also \cite{firpo2018synthetic} for extensions of SC placebo tests to non-constant propensity scores.
Here, we focus on the case where treatment happens at time $t$ and is assigned randomly to single unit $i$ with probability $p^{(t)}_i \in [0,1]$, where $\sumi p^{(t)}_i = 1$.
We can also accomodate the setting where the set of possibe time periods at which treatment can occur is a proper subset of $\{1,\ldots,T\}.$
We ask whether an estimator $\hat\tau^\gsc(\bu,\bv,\by,\momega)\equiv
 \sumituv\left\{M_{i0t}+ \sum_{j=1}^N M_{ijt} Y_{jt}\right\}$ is unbiased for $\tau(\bu,\bv) = \sumituv Y_{it}(1) -  Y_{it}(0)$ with respect to these propensity scores.
\begin{prop}
\label{prop:pMUSC}
    The estimator $\hat\tau^\gsc$ is unbiased for $\tau(\bu,\bv)$ (across values of potential outcomes) if and only if
    $
    \momega \in
        \calm^\musc_p=\biggl\{\momega\biggl| \sum_{j=1}^{N}M_{ijt}=0\: \forall i,t, \sum_{i=1}^N p^{(t)}_i M_{ijt}=0\: \forall j ,t\biggr\}.
    $
\end{prop}
Here, unbiasedness generalizes the adding-up condition $\sum_{i=1}^N M_{ijt}=0$ from the class of MUSC matrices to its propensity-weighted analogue $\sum_{i=1}^N p^{(t)}_i M_{ijt}=0$, which ensures that the bias is zero since
$
    \mme\left[\left.\hat\tau-\tau\right|\bv\right]=
    \sumt V_t \Big(\sumi \yitn\sumj p_j^{(t)} M_{jit} + \sumj p_j^{(t)} M_{j0t} \Big).
$
A natural analogue of the MUSC estimator is then
\begin{align}
\label{eqn:propensityscore}
    \momega^\musc_p(\by,\calm^\musc_p)\equiv\argmin_{\momega\in\calm^\musc_p} \sumi p^{(t)}_i \sumt\left\{\sum_{s < t} \left(M_{i0t}+\sumj M_{ijt} Y_{js}\right)^2\right\}.
\end{align}
Such an estimator could be used when treatment is assigned randomly.
Note that the variance estimator from \autoref{prop1} extends.
When the analyst has a choice over the treatment assignment, and $t=T$, the optimization in (\ref{eqn:propensityscore}) could also include the choice of propensity score.

We now illustrate how varying propensity scores can affect the Synthetic Control estimator, extending the motivating example in \autoref{motivating example}.
In this example, we consider the standard SC estimator, the USC estimator with equal propensities, and the USC estimator with non-constant propensities.
Recall that the standard SC estimator is biased under this design and the USC estimator corrects this bias by enforcing balance between the probability of being treated and being used as a control.
However, when the central unit is treated with higher probability of \sfrac{1}{2} (\autoref{fig:ABC} (b)), then the weight matrix that only uses the closest units as control in each case is the optimal unbiased solution.
In this specific example, this solution also coincides with the standard SC solution.

\begin{figure}
    \colorlet{unit1}{red!50!black}
    \colorlet{unit2}{green!50!black}
    \colorlet{unit3}{blue!50!black}

 \newcommand{\makecoordinates}{
            \draw[->] (1,0) -- (7,0) node[right]{$Y_{i1}$};
            \draw[color=unit1] (1 * 2,.2) -- (1 * 2,-.2) node[below]{AZ} ;
                \coordinate (1) at (1 * 2,0);
            \draw[color=unit2] (2 * 2,.2) -- (2 * 2,-.2) node[below]{CA} ;
                \coordinate (2) at (2 * 2,0);
                \draw[color=unit3] (3 * 2,.2) -- (3 * 2,-.2) node[below]{NY} ;
                \coordinate (3) at (3 * 2,0);
            \coordinate (offset) at (0,0.2);
            \coordinate (0) at (1.2,0);

    }

    \begin{center}                
    \begin{subfigure}[b]{0.4\textwidth}
            \centering
        \begin{tikzpicture}
            \makecoordinates{}

            \draw (0)+(offset) node[above] {\scriptsize $p_i=$};
            \draw (1)+(offset) node[above] {\sfrac{1}{3}};
            \draw (2)+(offset) node[above] {\sfrac{1}{3}};
            \draw (3)+(offset) node[above] {\sfrac{1}{3}};
            
            \draw[-{Latex[scale=2]},color=unit1] (1) to[bend right] node[below]{\scriptsize \sfrac{1}{2}} (2);
            \draw[-{Latex[scale=2]},color=unit3] (3) to[bend right] node[above]{\scriptsize \sfrac{1}{2}} (2);
            
            \draw[-{Latex[scale=2]},color=unit2] (2) to[bend right] node[above]{\scriptsize \sfrac{1}{2}} (1);
            \draw[-{Latex[scale=2]},color=unit3] (3) to[bend right=55] node[above]{\scriptsize \sfrac{1}{2}} (1);
            
            \draw[-{Latex[scale=2]},color=unit2] (2) to[bend right] node[below]{\scriptsize \sfrac{1}{2}} (3);
            \draw[-{Latex[scale=2]},color=unit1] (1) to[bend right=55] node[below]{\scriptsize \sfrac{1}{2}} (3);
            
            \coordinate (FIRST NE) at (current bounding box.north east);
            \coordinate (FIRST SW) at (current bounding box.south west);
            
        \end{tikzpicture}
        \caption{USC, constant propensity}
    \end{subfigure}
    \qquad
    \begin{subfigure}[b]{0.4\textwidth}
        \centering

        \begin{tikzpicture}
            \useasboundingbox (FIRST SW) rectangle (FIRST NE);
        
            \makecoordinates{}
            
            \draw (0)+(offset) node[above] {\scriptsize $p_i=$};
            \draw (1)+(offset) node[above] {\sfrac{1}{4}};
            \draw (2)+(offset) node[above] {\sfrac{1}{2}};
            \draw (3)+(offset) node[above] {\sfrac{1}{4}};
            
            \draw[-{Latex[scale=2]},color=unit2] (2) to[bend right] node[above]{\scriptsize 1} (1);
            \draw[-{Latex[scale=2]},color=unit2] (2) to[bend right] node[below]{\scriptsize 1} (3);
            
            \draw[-{Latex[scale=2]},color=unit1] (1) to[bend right] node[below]{\scriptsize \sfrac{1}{2}} (2);
            \draw[-{Latex[scale=2]},color=unit3] (3) to[bend right] node[above]{\scriptsize \sfrac{1}{2}} (2);
        \end{tikzpicture}
        \caption{USC, varying propensity}
    \end{subfigure}
    \end{center}
    
    \caption{Pre-treatment outcome in three-unit, two-period example with varying treatment propensities. An outgoing arrow represents the weight assigned to that unit when the target of the arrow is treated, and arrows are colored by the unit the respective weight is put on.}
    \label{fig:ABC}
\end{figure}

In this example, there is a set of propensity scores for which the standard SC estimator is unbiased.
This is not a coincidence.
As the following proposition shows, for the weights of every SC-type estimator there is a set of propensity scores such that the corresponding estimator is unbiased.

\begin{prop}
\label{prop:pexistence}
    For every SC weight matrix 
    \begin{align*}
        \momega \in \calm^{\ssc}=\left\{\momega\left|
        M_{iit}{=}1,\forall i,t;
         M_{ijt}{\leq} 0, \forall i, j{\geq} 1,t;
         \sum_{j=1}^{N}M_{ijt}=0\: \forall i,t;
         M_{i0t} {=} 0 \forall i
         \right.
        \right\}
    \end{align*}
    there exists a propensity-score vector $\mathbf{p}^{(t)} \in [0,1]^N, \sum_{i=1}^N p^{(t)}_i =1$ such that the corresponding estimator is unbiased with respect to $\mathbf{p}^{(t)}$ at treatment time $t$.
\end{prop}

This proposition does not rely on the weight matrix $\momega$ being the result of a specific optimization program, as it applies to any weight matrix that follows the basic structure of the SC matrices (without intercept).

To understand the propensity scores $\mathbf{p}^{(t)}$ that make the SC estimator associated with the weights $\momega$ unbiased,
we note that they can be interpreted as a measure of centrality in the network associated with $\momega$ in a precise way, where more central units correspond to higher propensity scores.
Specifically, let $\mathbf{W} = (W_{ij})_{i,j \in \{1,\ldots,N\}}$ be the edge flow matrix from \autoref{sec: network} corresponding to $\momega$ for treatment time $t$, meaning that $W_{ij} = - M_{jit}$ for $i \neq j$ and $W_{ii} = 0$.
Then the eigenvector centralities of vertices in this network are equivalent (up to normalization) to the propensity scores that ensure unbiasedness, where we consider the case where both are unique.

\begin{prop}
\label{prop:centrality}
    Assume that the network $\mathbf{W}$ associated with $\momega$ for treatment at $t$ is strongly connected (equivalently, that $\mathbf{W}$ is irreducible).
    Then the propensity score vector $\mathbf{p}^{(t)}$ for which the estimator is unbiased at $t$ is unique and the same as  the eigenvector centrality in the network $\mathbf{W}$ (with appropriate normalization).
\end{prop}

This connection between eigenvectors and unbiased propensities follows naturally from the representation of the estimator in terms of its (weighted) network adjacency matrix $\mathbf{W}$.
Writing $\mathbf{M}^{(t)} = (M_{ijt})_{i,j \in \{1,\ldots,N\}}$ for the $N \times N$ matrix corresponding to the SC weights when treatment happens at $t$,
the unbiasedness condition corresponds to $(\mathbf{M}^{(t)})' \mathbf{p}^{(t)} = \mathbf{0}$.
Since $(\mathbf{M}^{(t)})' = \mathbb{I} - \mathbf{W}$, $\mathbf{p}^{(t)}$ is an eigenvector of $\mathbf{W}$ with eigenvalue 1, which is also the largest eigenvalue and corresponds to the unique non-negative eigenvector if the network is strongly connected.
When the network is not strongly connected, we may still obtain a similar result for its components.

These results suggest ways in which considering varying propensity scores can be helpful when analyzing SC-type estimators.
First, when treatment is randomized according to a known probability distribution, then those probabilities affect the optimal USC and MUSC weights.
Second, the propensities that make an estimator unbiased have an intuitive interpretation as the eigenvector centralities of the network corresponding to the weight matrix of an SC estimator.
Third, even in the observational case, varying propensities could be used when some units can be considered to be more likely to receive treatment or to be more appropriate as controls, replacing binary inclusion criteria by treatment propensities.
Finally, when we choose propensity scores in the design of an experiment and plan to use an SC-type estimator, we can optimize the choice of propensities based on past outcomes to be better suited to their relationship, assigning more central units higher probabilities.

\section{AN ILLUSTRATION AND SOME SIMULATIONS}\label{section:illustration}

In this section we illustrate some of the methods proposed in the first part of this article. 
We first report the results of a simulation study based on real data. Second, we report results of a re-analysis of the 
 California smoking study \citep{abadie2010synthetic}.

\subsection{A Simulation Study}
\label{section:simulations}

We  perform a small simulation study to assess the properties of the MUSC estimator. Following \cite{Bertrand2004did} and \cite{arkhangelsky2019synthetic}, we use data from the Current Population Survey for $N=50$ states and $T=40$ years. The variables we analyze include state/year average log wages, hours, and the state/year unemployment rate. The true treatment effects are all zero by construction.  This allows us to calculate the RMSE. For each of the variables, we estimate the treatment effects using the Difference-in-Means (DiM) estimator, the standard Synthetic Control (SC) estimator, the Difference-in-Differences (DiD) estimator, and the Modified Unbiased Synthetic Control (MUSC) estimator. We also include for comparison the LASSO estimator based on regressing the period $T$ outcomes on all the lagged outcomes.
The LASSO estimator is not an SC-type estimator and is generally biased in our framework, and we merely include it as a benchmark that uses more information than the DiM estimator.
For the LASSO estimator, we choose the regularization parameter using leave-one-out cross validation. 

The first simulation study we conduct compares the performance of the six estimators in terms of RMSE. The study is designed as follows. For each treated period $T\in\{21,\ldots,40\}$, we use $T-1$ pre-treatment periods to estimate weights for the six estimators and discard all data after $T$. Then, iterating through all 50 states, we pretend that each state has been selected for treatment and calculate the corresponding estimated treatment effects. Lastly, we average over all states to summarize the performance for a single treated period.

In \autoref{tab:rmseaverage1}, we report the results averaged over all twenty years. 
We report the results for the setting with 50 units, as well as for settings with 10 and 5 units to assess the relative performance with fewer cross-sectional units.
We find that the RMSE is substantially lower for the SC and MUSC estimator compared to the DiM estimator for all variables. In Table 1 in the appendix, we document that these results hold across years. The SC and MUSC estimators perform comparably to the LASSO estimator for log wages and outperform it for the other two variables in the case with $N=50$, with the DiD estimator performing worse than any of these three. Note that the SC and MUSC estimators have similar RMSE in this case. In the cases with $N=10$ and $N=5$, the MUSC estimator substantially outperforms the other methods except for the DiD estimator, with the latter performing slightly worse for $N=10$ and slightly better for $N=5$. Overall, the MUSC estimator performs consistently well. Unsurprisingly the relative performance of the LASSO  estimator deteriorates sharply when the number of units is small and there are not enough control units to estimate the LASSO parameters well. Additional simulations show  that the DiD estimator performs relatively poorly when many of the units  are well approximated by a small number of control units.

\begin{table}[!htbp]
    \centering
    \caption{Simulation Experiment Based on CPS Data Averaged over States and Years -- Roor-Mean-Squared-Error (RMSE)}
    \label{tab:rmseaverage1}
    \begin{threeparttable}
        \small
    \begin{tabular}{lccccccccc}
    \toprule
    & \multicolumn{3}{c}{$N=50$} & \multicolumn{3}{c}{$N=10$} & \multicolumn{3}{c}{$N=5$}\\
    \cmidrule(lr){2-4} \cmidrule(lr){5-7} \cmidrule(lr){8-10}
          Est & Ln(Wage) & Hours & U-Rate & Ln(Wage) & Hours & U-Rate & Ln(Wage) & Hours & U-Rate\\ \midrule
         DiM & 0.105 & 1.197 & 0.015 & 0.122& 1.558 & 0.016 & 0.144 & 1.380 & 0.015\\
          SC & 0.051 & 0.918 & 0.013& 0.070 & 1.045 & 0.015 & 0.079 & 1.010 & 0.014\\
                SC--NR & 0.076 & 1.350 & 0.019 &  0.064& 1.049 & 0.017 & 0.058 & 0.965 & 0.013\\
    MUSC & 0.053 & 0.903 & 0.013& 0.061 & 0.960 & 0.015 &0.057 & 0.944 & 0.013\\
          LASSO & 0.051 & 0.952 & 0.013 & 0.077 & 1.313 & 0.017 & 0.116 & 1.251 & 0.015\\
          DiD & 0.063 & 0.976 & 0.013 & 0.066 & 0.982 & 0.014 & 0.055 & 0.929 & 0.013 \\\bottomrule
    \end{tabular}
     {\footnotesize DiM: Difference-in-Means estimator, SC: Synthetic Control estimator of
\cite{abadie2010synthetic}, SC--NR: Synthetic Control Estimator with no adding up restriction on the weights, MUSC: Modified Unbiased Synthetic Control estimator, DiD: Difference-in-Differences estimator.}
\end{threeparttable}
\end{table}

The second simulation study demonstrates the properties of our proposed unbiased variance estimator and the placebo variance estimator. Here we focus on average log wages and fix $T=40$ as the treated period. Moreover, we decrease the sample to $N=20$ units in total. \autoref{tab: vars} reports standard errors based on the true variance along with the estimates (averaged over all units) based on our variance estimator and using the placebo approach. We find that our estimator is indeed unbiased. The placebo approach is very modestly biased -- the direction of the bias depends on which estimator is used. For the DiM, the DiD, and the SC estimator, the placebo estimator is upward biased; for the MUSC estimator it is downward biased.

\begin{table}[h!]
    \caption{Simulation Experiment Based on CPS Data (Log Wages) by State and Year for $N=20$ units and Treated Period $T=40$ -- Average Standard Error}
    \label{tab: vars}
    \centering
    \begin{tabular}{lccc}
         \toprule
          & $\sqrt{{V}_{\textnormal{true}}}$ &  $\sqrt{\hat{V}_{\textnormal{GSC}}}$ & $\sqrt{\hat{V}_{\textnormal{PCB}}}$ \\ 
          \midrule
          DiM & 0.1157 & 0.1157 & 0.1158\\
          SC & 0.0518 & 0.0518 & 0.0521\\
          MUSC & 0.0490 & 0.0490 & 0.0489\\
          DiD & 0.0553 & 0.0553 & 0.0554\\ \bottomrule
    \end{tabular} \\ \vskip0.2cm
{\footnotesize DiM: Difference in Means, SC: Synthetic Control, MUSC: Modified Synthetic Control, DiD: Difference in Differences, GSC: General Synthetic Control, PCB: Placebo}
\end{table}

A third simulation exercise study illustrates the coverage and length of randomization-based confidence intervals as well as confidence intervals based on Normal-distribution approximation using our unbiased variance estimate. We discuss the construction of the randomization based confidence intervals in \autoref{subsec:randomization}. \autoref{tab: randomization cis sims 5000 ordering} shows the results. We find that our randomization-based confidence intervals provide correct coverage, while Normality-based intervals may under- or over-cover. This is unsurprising because they are not formally justified in the case with a single treated unit/time period combination, which we focus on in this article.

\begin{table}[ht!]
    \centering 
    
    \caption{Confidence Intervals}
    \small
    \begin{tabular}{llcccc}

    \toprule
        & & \multicolumn{2}{c}{\textbf{Randomization-Based}} & \multicolumn{2}{c}{\textbf{Normality-Based}} \\
            \cmidrule(lr){3-4}
\cmidrule(lr){5-6}
       {Variable} &  {Estimator} &  {Coverage } & {Length} & {Coverage} & {Length} \\\midrule
        Urate & {DiM} & 0.950   & 0.061  &  0.947  &  0.059\\
        & {SC} & 0.950  &  0.055  &  0.944  &  0.051\\
        & {MUSC} & 0.950  &  0.056  &  0.945  &  0.051\\
        \midrule
        Hours & {DiM} &  0.950  &  5.051  &  0.943  &  4.693\\
         & {SC} &0.950  &  3.905  &  0.951  &  3.597\\
        & {MUSC} & 0.950  &  3.996  &  0.939  &  3.539\\
        \midrule
        Lwage &{DiM}& 0.950  &  0.429  &  0.951  &  0.410\\
         & {SC} & 0.950  &  0.219  &  0.939  &  0.200\\
        & {MUSC} & 0.950  &  0.238  &  0.945  &  0.207\\
        \bottomrule
    \end{tabular}
    
    \medskip
        Coverage and length of confidence intervals based on the MUSC estimator and, in the case of Normality-based intervals, our unbiased estimator of its variance. Coverage and length is averaged over 5,000 random draws of the treated unit and period from $T = 21 \cdot \cdot \cdot 40$, $S = 5,000$.
    \label{tab: randomization cis sims 5000 ordering}
\end{table}

\subsection{The California Smoking Study}

\begin{figure}
    \centering
    \includegraphics[width = 0.75\linewidth]{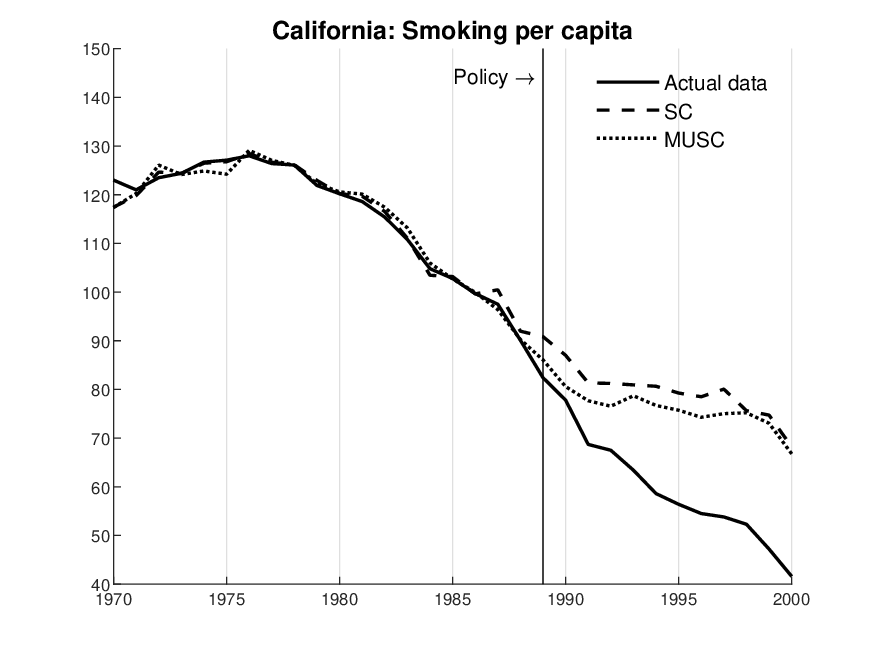}
    \caption{Pre- and Post-Treatment fit of SC and MUSC}
    \label{fig:smoke}
\end{figure}

Next, we  turn to the data from the California smoking study \citep{abadie2010synthetic}. In \autoref{fig:smoke} we compare the SC and MUSC estimates.
We find that the pre-treatment fit is similar for both estimators despite the additional restriction. In addition, the point estimates are similar. The interpretation is that the number of ``similar'' control units is large enough that a single additional restriction (and the relaxing of another one) does not affect the goodness of fit substantially.

\section{CONCLUSION}

In this article, we study Synthetic Control (SC) methods from a design perspective. We show that when a randomized experiment is conducted, the standard SC estimator is biased. However, a minor modification of the SC estimator is unbiased under randomization, and in cases with few treated units can have RMSE properties superior to those of the standard Difference-in-Means estimator. We show that the design perspective also has implications for observational studies. We propose a variance estimator validated by randomization.

In the online appendix we discuss some extensions, including the results for the case with multiple treated units in Appendix D.1 and the case where the estimand is the average effect for all units in the treated period, $\tauv$, in Appendix D.2.

\section*{ACKNOWLEDGEMENTS}
    This research was generously supported by ONR grants N00014-17-1-2131 and N00014-19-1-2468. The authors would like to thank Alberto Abadie, Kirill Borusyak, Jonathan Roth, Jeffrey Wooldridge, the editor, the associate editor, three anonymous referees, and seminar audiences at the NBER and at USC for helpful comments and general discussions on the topic of this article. The authors report there are no competing interests to declare.

\section*{SUPPLEMENTARY MATERIAL}
Supplementary materials are four appendices. Appendix A: Additional table. Appendix B: Additional theoretical results. Appendix C: Proofs. Appendix D: Extensions and Generalizations.

\bibliographystyle{apalike}
\bibliography{references}

\begin{thebibliography}{}

\bibitem[Abadie, 2021]{abadie2019using}
Abadie, A. (2021).
\newblock Using synthetic controls: Feasibility, data requirements, and methodological aspects.
\newblock {\em Journal of Economic Literature}, 59(2):391--425.

\bibitem[Abadie et~al., 2020]{abadie2020sampling}
Abadie, A., Athey, S., Imbens, G.~W., and Wooldridge, J.~M. (2020).
\newblock Sampling-based versus design-based uncertainty in regression analysis.
\newblock {\em Econometrica}, 88(1):265--296.

\bibitem[Abadie et~al., 2010]{abadie2010synthetic}
Abadie, A., Diamond, A., and Hainmueller, J. (2010).
\newblock Synthetic control methods for comparative case studies: Estimating the effect of california's tobacco control program.
\newblock {\em Journal of the American statistical Association}, 105(490):493--505.

\bibitem[Abadie et~al., 2015]{abadie2014}
Abadie, A., Diamond, A., and Hainmueller, J. (2015).
\newblock Comparative politics and the synthetic control method.
\newblock {\em American Journal of Political Science}, pages 495--510.

\bibitem[Abadie and Gardeazabal, 2003]{abadie2003}
Abadie, A. and Gardeazabal, J. (2003).
\newblock The economic costs of conflict: A case study of the basque country.
\newblock {\em American Economic Review}, 93(-):113--132.

\bibitem[Abadie and Imbens, 2006]{abadie2006}
Abadie, A. and Imbens, G.~W. (2006).
\newblock Large sample properties of matching estimators for average treatment effects.
\newblock {\em Econometrica}, 74(1):235--267.

\bibitem[Abadie and L’Hour, 2017]{abadie2017penalized}
Abadie, A. and L’Hour, J. (2017).
\newblock A penalized synthetic control estimator for disaggregated data.
\newblock {\em Work. Pap., Mass. Inst. Technol., Cambridge, MA}.

\bibitem[Abadie and Zhao, 2021]{abadie2021synthetic}
Abadie, A. and Zhao, J. (2021).
\newblock Synthetic controls for experimental design.
\newblock {\em arXiv preprint arXiv:2108.02196}.

\bibitem[Amjad et~al., 2018]{amjad2018robust}
Amjad, M., Shah, D., and Shen, D. (2018).
\newblock Robust synthetic control.
\newblock {\em The Journal of Machine Learning Research}, 19(1):802--852.

\bibitem[Arkhangelsky et~al., 2019]{arkhangelsky2019synthetic}
Arkhangelsky, D., Athey, S., Hirshberg, D.~A., Imbens, G.~W., and Wager, S. (2019).
\newblock Synthetic difference in differences.
\newblock Technical report, National Bureau of Economic Research.

\bibitem[Athey et~al., 2021]{athey2017matrix}
Athey, S., Bayati, M., Doudchenko, N., Imbens, G., and Khosravi, K. (2021).
\newblock Matrix completion methods for causal panel data models.
\newblock {\em Journal of the American Statistical Association}.

\bibitem[Athey and Imbens, 2018]{athey2018design}
Athey, S. and Imbens, G.~W. (2018).
\newblock Design-based analysis in difference-in-differences settings with staggered adoption.
\newblock Technical report, National Bureau of Economic Research.

\bibitem[Ben-Michael et~al., 2020]{ben2018augmented}
Ben-Michael, E., Feller, A., and Rothstein, J. (2020).
\newblock The augmented synthetic control method.
\newblock {\em arXiv preprint arXiv:1811.04170}.

\bibitem[Berman and Plemmons, 1994]{Berman1994-ph}
Berman, A. and Plemmons, R.~J. (1994).
\newblock {2. Nonnegative Matrices}.
\newblock In {\em {Nonnegative Matrices in the Mathematical Sciences}}, Classics in Applied Mathematics, pages 26--62. Society for Industrial and Applied Mathematics.

\bibitem[Bertrand et~al., 2004]{Bertrand2004did}
Bertrand, M., Duflo, E., and Mullainathan, S. (2004).
\newblock How much should we trust differences-in-differences estimates?
\newblock {\em The Quarterly Journal of Economics}, 119(1):249--275.

\bibitem[Cavallo et~al., 2013]{cavallo2013catastrophic}
Cavallo, E., Galiani, S., Noy, I., and Pantano, J. (2013).
\newblock Catastrophic natural disasters and economic growth.
\newblock {\em Review of Economics and Statistics}, 95(5):1549--1561.

\bibitem[Chen, 2022]{chen2022synthetic}
Chen, J. (2022).
\newblock Synthetic control as online linear regression.
\newblock {\em arXiv preprint arXiv:2202.08426}.

\bibitem[Chernozhukov et~al., 2017]{chernozhukov2017exact}
Chernozhukov, V., Wuthrich, K., and Zhu, Y. (2017).
\newblock An exact and robust conformal inference method for counterfactual and synthetic controls.
\newblock {\em arXiv preprint arXiv:1712.09089}.

\bibitem[Coffman and Noy, 2012]{coffman2012hurricane}
Coffman, M. and Noy, I. (2012).
\newblock Hurricane iniki: measuring the long-term economic impact of a natural disaster using synthetic control.
\newblock {\em Environment and Development Economics}, 17(2):187--205.

\bibitem[Cunningham, 2018]{cunningham2018causal}
Cunningham, S. (2018).
\newblock {\em Causal inference: The mixtape}.
\newblock Yale University Press.

\bibitem[de~Paula, 2020]{De_Paula2020-ux}
de~Paula, {\'A}. (2020).
\newblock {Econometric Models of Network Formation}.
\newblock {\em Annual review of economics}, 12(1):775--799.

\bibitem[Doudchenko and Imbens, 2016]{doudchenko2016balancing}
Doudchenko, N. and Imbens, G.~W. (2016).
\newblock Balancing, regression, difference-in-differences and synthetic control methods: A synthesis.
\newblock Technical report, NBER.

\bibitem[Ferman and Pinto, 2017]{ferman2017placebo}
Ferman, B. and Pinto, C. (2017).
\newblock Placebo tests for synthetic controls.

\bibitem[Ferman and Pinto, 2019]{ferman2019synthetic}
Ferman, B. and Pinto, C. (2019).
\newblock Synthetic controls with imperfect pre-treatment fit.
\newblock {\em arXiv preprint arXiv:1911.08521}.

\bibitem[Firpo and Possebom, 2018]{firpo2018synthetic}
Firpo, S. and Possebom, V. (2018).
\newblock Synthetic control method: Inference, sensitivity analysis and confidence sets.
\newblock {\em Journal of Causal Inference}, 6(2).

\bibitem[Fisher, 1937]{fisher1937design}
Fisher, R.~A. (1937).
\newblock {\em The design of experiments}.
\newblock Oliver And Boyd; Edinburgh; London.

\bibitem[Freedman, 2008]{freedman2008regression}
Freedman, D.~A. (2008).
\newblock On regression adjustments in experiments with several treatments.
\newblock {\em The annals of applied statistics}, 2(1):176--196.

\bibitem[Hahn and Shi, 2016]{hahn2016}
Hahn, J. and Shi, R. (2016).
\newblock Synthetic control and inference.
\newblock {\em Available at UCLA}.

\bibitem[Imbens and Rubin, 2015]{imbens2015causal}
Imbens, G.~W. and Rubin, D.~B. (2015).
\newblock {\em Causal Inference in Statistics, Social, and Biomedical Sciences}.
\newblock Cambridge University Press.

\bibitem[Jackson, 2010]{Jackson2010-tw}
Jackson, M.~O. (2010).
\newblock {\em {Social and Economic Networks}}.
\newblock Princeton University Press.

\bibitem[Lei and Cand{\`e}s, 2020]{lei2020conformal}
Lei, L. and Cand{\`e}s, E.~J. (2020).
\newblock Conformal inference of counterfactuals and individual treatment effects.
\newblock {\em arXiv preprint arXiv:2006.06138}.

\bibitem[Li, 2020]{li2020}
Li, K.~T. (2020).
\newblock Statistical inference for average treatment effects estimated by synthetic control methods.
\newblock {\em Journal of the American Statistical Association}, 115(532):2068--2083.

\bibitem[Lin, 2013]{lin}
Lin, W. (2013).
\newblock Agnostic notes on regression adjustments for experimental data: Reexamining freedman's critique.
\newblock {\em The Annals of Applied Statistics}, 7(1):295--318.

\bibitem[Liu, 2015]{liu2015spillovers}
Liu, S. (2015).
\newblock Spillovers from universities: Evidence from the land-grant program.
\newblock {\em Journal of Urban Economics}, 87:25--41.

\bibitem[Manski and Pepper, 2018]{manski2018right}
Manski, C.~F. and Pepper, J.~V. (2018).
\newblock How do right-to-carry laws affect crime rates? coping with ambiguity using bounded-variation assumptions.
\newblock {\em Review of Economics and Statistics}, 100(2):232--244.

\bibitem[Neyman, 1990]{neyman1923}
Neyman, J. (1923/1990).
\newblock On the application of probability theory to agricultural experiments. essay on principles. section 9.
\newblock {\em Statistical Science}, 5(4):465--472.

\bibitem[Rambachan and Roth, 2020]{Rambachan2020-ob}
Rambachan, A. and Roth, J. (2020).
\newblock {Design-Based Uncertainty for Quasi-Experiments}.
\newblock {\em arXiv preprint arXiv:2008.00602}.

\bibitem[Rosenbaum, 2002]{rosenbaum_book}
Rosenbaum, P.~R. (2002).
\newblock {\em Observational Studies}.
\newblock Springer.

\bibitem[Roth and Sant'Anna, 2021]{roth2021efficient}
Roth, J. and Sant'Anna, P.~H. (2021).
\newblock Efficient estimation for staggered rollout designs.
\newblock {\em arXiv preprint arXiv:2102.01291}.

\bibitem[Rubin, 1974]{rubin1974estimating}
Rubin, D.~B. (1974).
\newblock Estimating causal effects of treatments in randomized and nonrandomized studies.
\newblock {\em Journal of educational Psychology}, 66(5):688.

\bibitem[Rubin, 2008]{rubin2008objective}
Rubin, D.~B. (2008).
\newblock For objective causal inference, design trumps analysis.
\newblock {\em The Annals of Applied Statistics}, pages 808--840.

\bibitem[Sekhon and Shem-Tov, 2020]{sekhon2020inference}
Sekhon, J.~S. and Shem-Tov, Y. (2020).
\newblock Inference on a new class of sample average treatment effects.
\newblock {\em Journal of the American Statistical Association}, pages 1--7.

\bibitem[Xu, 2017]{xu2017generalized}
Xu, Y. (2017).
\newblock Generalized synthetic control method: Causal inference with interactive fixed effects models.
\newblock {\em Political Analysis}, 25(1):57--76.

\end{thebibliography}

\newpage

\setcounter{page}{1}
\setcounter{section}{0}
\appendix
\section{ ADDITIONAL TABLES}
\begin{table}[!htbp]
\centering
\rotatebox{90}{
\begin{threeparttable}
\centering  
\setlength{\tabcolsep}{3pt}
\caption{Simulation Experiment Based on CPS Data by State and Year -- RMSE} 
\label{tbl:full}
\scriptsize
\begin{tabular}{@{\extracolsep{1pt}} ccccccccccccccccccc} \toprule Treated &  \multicolumn{6}{c}{\textbf{Log Wages}} & \multicolumn{6}{c}{\textbf{Hours}}& \multicolumn{6}{c}{\textbf{Unemployment Rate}}\\
\cmidrule(lr){2-7}
\cmidrule(lr){8-13}
\cmidrule(lr){14-19}
Period & \multicolumn{1}{c}{$\ddd$} & $\ssc$ & SC- NR & $\musc$  & LASSO & DiD & $\ddd$ & $\ssc$ & SC- NR  &  $\musc$ & LASSO& DiD& $\ddd$ & $\ssc$ & SC- NR  &  $\musc$ & LASSO& DiD\\ \midrule
$T = 21$ &\multicolumn{1}{|c} {0.1157}  &  0.0543 &  0.0646 &   0.0550 & 0.0560 & 0.0636 & 1.4838    & 0.9797& 1.3138 &  0.9340 &   1.1999 &  1.1063& 0.0123  &  0.0111&  0.0124  &  0.0102 &  0.0110& 0.0114\\ 
$T = 22$ & \multicolumn{1}{|c}{0.1112}  &  0.0496 & 0.0639 &   0.0502 & 0.0512 & 0.0708  &1.3764     & 0.8204& 1.2410 &  0.9170 & 0.9963 &  1.0276&  0.0107  &  0.0086 & 0.0100 &   0.0084 &  0.0102 & 0.0096\\
$T = 23$ & \multicolumn{1}{|c}{0.1089}  &  0.0435 & 0.0587 & 0.0484 & 0.0409 & 0.0540& 1.2821     & 0.9091& 0.9691 &  0.9009  &  0.9904  &0.9197& 0.0114  &  0.0111 & 0.0132 &   0.0116  &  0.0107 & 0.0110\\
$T = 24$ & \multicolumn{1}{|c}{0.1143}  &  0.0421 & 0.0596 &  0.0450& 0.0452& 0.0626 & 1.2400     & 0.9569& 0.9305 &  0.8829  & 0.9524& 0.9747& 0.0145  &  0.0146 & 0.0142 &   0.0142 & 0.0145 & 0.0148\\
$T = 25$ & \multicolumn{1}{|c}{0.1136 } &  0.0465 & 0.0523 &   0.0500 &  0.0446& 0.0602& 1.3125     & 0.9074& 1.0759 &  0.8896 &  0.9011& 1.0567&   0.0124  &  0.0110& 0.0157  &   0.0115 &  0.0125 & 0.0114\\
$T = 26$ &\multicolumn{1}{|c} {0.1111}  &  0.0487 & 0.0748  &   0.0471 & 0.0509 & 0.0639 & 1.1424     & 0.8509& 1.0917 &  0.8347 & 0.8456&  0.8544& 0.0132  &  0.0136& 0.0217  &   0.0146  & 0.0132 & 0.0123\\
$T = 27$ & \multicolumn{1}{|c}{0.1105}  &  0.0481 & 0.0565  &   0.0493 &  0.0439 & 0.0506 & 1.1926     & 0.8612& 1.2655 &  0.8141 & 0.9746& 0.9063&  0.0155  &  0.0145&  0.0227 &   0.0149 &  0.0155& 0.0139\\
$T = 28$ & \multicolumn{1}{|c}{0.1014}  &  0.0551 & 0.0910 &  0.0674 &  0.0597 & 0.0640 & 1.0627     & 0.7611& 1.1673 &  0.7660 &  0.8451& 0.9322&  0.0136  &  0.0105&  0.0153 &  0.0106&  0.0122 & 0.0099\\
$T = 29$ & \multicolumn{1}{|c}{0.0964}  &  0.0473 & 0.0782 &  0.0594 &  0.0487 & 0.0628 & 1.1342     & 0.8552& 0.8681 &  0.8686 &  0.8797& 1.0101&  0.0137  &  0.0124&  0.0175 &   0.0118 &  0.0133 & 0.0109\\
$T = 30$ & \multicolumn{1}{|c} {0.0980} &   0.0505 &  0.0661 &    0.0516& 0.0562 & 0.0626 & 1.0741     & 0.7765& 0.9716 &  0.7755 &  0.7318& 0.8350&  0.0137  &  0.0141& 0.0201  &  0.0147 &  0.0137 & 0.0140\\
$T = 31$ & \multicolumn{1}{|c} {0.1084} &   0.0410 &0.0576 &   0.0393 &  0.0429 & 0.0691 & 1.2340     & 0.9700& 1.4479 & 0.9537 &  0.9405& 1.0748&  0.0209  &  0.0180& 0.0260  &   0.0175 &  0.0170 & 0.0196\\
$T = 32$ &\multicolumn{1}{|c} {0.1049} &   0.0460 & 0.0746 &   0.0468 &  0.0475 & 0.0600 & 1.0571     & 0.8496& 1.1947 &  0.8209&  0.8495&0.8910&  0.0214  &  0.0182&  0.0266 &   0.0167 & 0.0180 & 0.0177\\
$T = 33$ &\multicolumn{1}{|c} {0.1046} &   0.0530 & 0.0950 &   0.0536 & 0.0485 & 0.0594 & 1.2947     & 1.0018& 1.8156 &  0.9696 &  1.0183& 1.0211&  0.0215  &  0.0161 & 0.0185 &   0.0151 & 0.0133 & 0.0175\\
$T = 34$ &\multicolumn{1}{|c} {0.1060} &   0.0625 & 0.0877 &   0.0662 &  0.0610 & 0.0741 & 1.1640     & 0.9174& 1.3004 &  0.9007 &   0.9836& 1.0024& 0.0182  &  0.0131& 0.0230  &    0.0129 &  0.0122 & 0.0142\\
$T = 35$ &\multicolumn{1}{|c} {0.1053}  &  0.0560 & 0.0981 &   0.0537 &  0.0529 & 0.0623 & 1.2233     & 1.0824& 1.7619 &  0.9914 & 1.0680&  0.9688& 0.0201  &  0.0137 & 0.0203 &    0.0137 &  0.0166 & 0.0179\\
$T = 36$ &\multicolumn{1}{|c} {0.0985}   & 0.0594 & 0.1145  &  0.0535 & 0.0530 & 0.0593 & 1.1267     & 0.8083& 1.0229 & 0.7396 & 0.8650& 0.9133& 0.0150  &  0.0124 &  0.0159 &    0.0119 &  0.0134 & 0.0121\\
$T = 37$ &\multicolumn{1}{|c} {0.1017}    &0.0605 & 0.0719 &   0.0579 & 0.0548 & 0.0672 & 1.3066     & 1.2080& 2.2323 &  1.1235 &  1.2517& 1.0831& 0.0156  &  0.0128  & 0.0259 &  0.0131 &  0.0145& 0.0126\\
$T = 38$ &\multicolumn{1}{|c} {0.0929}   & 0.0580 & 0.1007 &  0.0615 &  0.0607 & 0.0665 & 0.9917     & 0.7742& 1.5303 &  0.7929 &  0.7677& 0.8234& 0.0126  &  0.0116  &  0.0249 & 0.0117 & 0.0113 & 0.0108\\
$T = 39$ &\multicolumn{1}{|c} {0.0853}  &  0.0459 & 0.0892 &  0.0554 &0.0435 & 0.0628 & 0.8437     & 0.7979& 1.7969 &  0.9475  & 0.6663&  0.9233& 0.0112  &  0.0120  & 0.0205 &   0.0122 &  0.0112 & 0.0114 \\
$T = 40$ &\multicolumn{1}{|c} {0.1051}  &  0.0517 & 0.0634  &   0.0479 & 0.0527 & 0.0598 & 1.4048     & 1.2714& 2.0027 &  1.2382 &  1.3024& 1.1896&  0.0126  &  0.0112&  0.0188 &    0.0106 &  0.0114& 0.0103\\ \midrule
\textbf{Average} & \multicolumn{1}{|c}{0.1047}  &  0.0510&  0.0759 &  0.0530 & 0.0507 & 0.0628 & 1.1974     & 0.9180& 1.3500 &  0.9031 &   0.9515& 0.9757& 0.0150  &  0.0130 & 0.0192 &    0.0129  & 0.0132& 0.0132\\
\bottomrule\end{tabular}
     \begin{tablenotes} 
     \small
     \item DiM: Difference-in-Means estimator, SC: Synthetic Control estimator of \cite{abadie2010synthetic}, MUSC: Modified Unbiased Synthetic Control estimator,
     DiD: Difference-in-Differences estimator.
    \end{tablenotes}    
    \end{threeparttable}}
\end{table}

\newpage
\section{ADDITIONAL RESULTS}

\subsection{Placebo Variance Examples}

Here we present examples for which the placebo variance is biased. For simplicity, we consider the case where treatment occurs in the last of $T=3$ periods and there are $N=4$ units.
In this setting, the placebo variance can be biased downward:

\begin{example}
Suppose that for some arbitrary $a,b,c,d$,  
\[
 \by(0)=
\begin{psmallmatrix}
a& b&   0  \\
a& b&    1\\
c& d&    0\\
c& d&    1
\end{psmallmatrix},\ \ 
{\rm leading\ to}\ \ \momega^\musc=
\begin{psmallmatrix}
0&    1 & -1 & 0  & 0 \\
0&     -1 & 1 & 0 & 0\\
0&     0 & 0 & 1 & -1\\
0&     0 & 0 & -1 & 1
\end{psmallmatrix},
\]
so that the units are matched in pairs. If unit $1$ is treated, the estimator is $Y_{13}-Y_{23}$, with error $ Y_{13}(0)-Y_{23}(0)=-1$. Similar calculations for the other three units show that the squared error is always equal to 1, and hence 
the true variance is 1.

Now let us calculate the placebo variance. Here we exploit the fact that with three units the weights for all units are equal. This leads to 
\[ \momega^{(1)}=
\begin{psmallmatrix}
0&    0 & 0 & 0  & 0 \\
-((a+b)-(c+d))/2&   0 & 1 & -1/2 & -1/2\\
-((c+d)-(a+b))/4&    0 & -1/2 & 1 & -1/2\\
-((c+d)-(a+b))/4&     0 & -1/2 & -1/2 & 1
\end{psmallmatrix},
\]
Then the placebo variance is smaller in expectation than the true variance.
\end{example}

In the same setting, the placebo variance can be biased upward:

\begin{example}
 Suppose that
\[ \momega=
\begin{psmallmatrix}
     1 & -1 & 0  & 0 \\
     -1 & 1 & 0 & 0\\
     0 & 0 & 1 & -1\\
     0 & 0 & -1 & 1
\end{psmallmatrix}
,
\hskip1cm
 \by_{\cdot T}=
\begin{psmallmatrix}
    1  \\
     1\\
     0\\
     0
\end{psmallmatrix}
,
\]
so the units are matched in pairs, and the matching is of perfect quality.
Then the placebo variance is higher in expectation than the true variance.
\end{example}

\subsection{Non-Stochastic Weights}

Here, we argue formally that we can consider the weights to be non-stochastic for the main analysis in our article. 
For a given set $\calm$ we can
write each estimator as
\begin{equation}\label{sc_sets}
	 \hat\tau(\bu,\bv,\by,\momega(\by,\calm))=
	\sumituv\left\{M_{i0t}(\by,\calm)+ \sum_{j=1}^N  M_{ijt}(\by,\calm) Y_{jt}\right\}.
	\end{equation}
For this class of estimators we can view the weights as non-stochastic:	
\begin{lemma} For all $\byn$, $\bye$, $\bu$,
$\bv$, and $\calm$,
\begin{equation}
\hat\tau(\bu,\bv,\by,\momega(\by,\calm))=
\hat\tau(\bu,\bv,\by,\momega(\byn,\calm))
\end{equation}
\end{lemma}	
This representation is useful because the properties of 
$\hat\tau(\bu,\bv,\by,\momega(\byn,\calm))$ are easier to establish under assumptions on $\bv$ and $\bu$ than those of 
$\hat\tau(\bu,\bv,\by,\momega(\by,\calm))$ for the general case.

\section{PROOFS}

\begin{proof}[Proof of Proposition 1]
We first consider the case without intercept.
As a preliminary calculation, note that for $k,j,j' \in \{1,\ldots,n\}$
\begin{align}
\label{eqn:balancedcounting}
    \sum_{i=1}^N
    \sum_{k,j,j' \neq i}
    \frac{1}{N - |\{k,j,j'\}|} a_{kjj'}
    =
    \sum_{k,j,j'} a_{kjj'},
\end{align}
since every term $kjj'$ term appears $N - |\{k,j,j'\}|$ times in the sum on the left.
Let now
\[
    a_{kjj'} = M_{kj} (Y_j(0) - Y_k(0)) \cdot M_{kj'} (Y_{j'}(0) - Y_{k}(0)),
\]
where for simplicity we fix the period $t$, drop all time indices to set $M_{ij}=M_{ijt}$, and write $\hat{\mmv}_i$ for the variance estimator when $U_i=1$.
Then $a_{kjj'} = 0$ for $k \in \{j,j'\}$ and thus
\[
    \frac{1}{N} \sum_{i=1}^N \underbrace{\left(\frac{1}{N{-}3} \sum_{k \neq i} \left(\sum_{j \neq i} M_{kj} (Y_j(0) - Y_k(0))\right)^2 - \frac{1}{(N{-}3)(N{-}2)} \sum_{k, j \neq i} M^2_{kj} (Y_j(0) - Y_k(0))^2
    \right)}_{=\hat{\mmv}_i}
    \]
    \[\hskip1cm
=
    \frac{1}{N}
    \sum_{i=1}^N
    \Big(
        \sum_{k,j,j'\neq i} \frac{1}{N-3} a_{kjj'} - \sum_{\substack{k,j,j'\neq i \\ j = j'}} \underbrace{\frac{1}{(N-3)(N-2)}}_{ = \frac{1}{N-3} - \frac{1}{N-2}} a_{kjj'}
    \Big)
    \]
    \[\hskip1cm
    =
    \frac{1}{N}
    \Big(
    \sum_{i=1}^N
        \sum_{\substack{k,j,j'\neq i \\ |\{k,j,j'\}|=3}} \frac{1}{N-3} a_{kjj'} 
        +
        \sum_{\substack{k,j,j'\neq i \\ |\{k,j,j'\}|=2}} \frac{1}{N-2} \underbrace{a_{kjj'}}_{=0 \text{ for $j \neq j'$}}
        +
        \sum_{\substack{k,j,j'\neq i \\ |\{k,j,j'\}|=1}} \frac{1}{N-1} \underbrace{a_{kjj'}}_{=0}
    \Big)
    \]
    \[\hskip1cm
    =
    \frac{1}{N}\sum_{i=1}^N
    \sum_{k,j,j' \neq i}
    \frac{1}{N - |\{k,j,j'\}|} a_{kjj'}
    \stackrel{\text{(\ref{eqn:balancedcounting})}}{=}
    \frac{1}{N}\sum_{k,j,j'} a_{kjj'}
    \]
    \[\hskip1cm
    =
    \frac{1}{N} \sum_{i=1}^N \left( \sum_{j=1}^N M_{ij} (Y_j(0) - Y_i(0))\right)^2
    =
    \frac{1}{N} \sum_{i=1}^N \left( \sum_{j=1}^N M_{ij} Y_j(0)\right)^2
    =
    \mmv.
    \]
Here, we have used that $\sum_{j=1}^N M_{ij} = 0$.

With an intercept we note that
\begin{align*}
    \mmv
    &=
    \frac{1}{N} \sum_{i=1}^N \left( M_{i0} + \sum_{j=1}^N M_{ij} Y_j(0)\right)^2
    =
    \frac{1}{N} \sum_{i=1}^N \left( M_{i0} + \sum_{j=1}^N M_{ij} (Y_j(0) - Y_i(0))\right)^2
    \\
    &=
    \frac{1}{N} \sum_{i=1}^N M^2_{i0}
    +
    \frac{2}{N} \sum_{i=1}^N
    M_{i0} \left( \sum_{j=1}^N M_{ij} (Y_j(0) - Y_i(0))\right)
    +
    \frac{1}{N} \left( \sum_{j=1}^N M_{ij} (Y_j(0) - Y_i(0))\right)^2,
\end{align*}
where
\begin{align*}
    \frac{2}{N-2} \sum_{k \neq i} M_{k0} \left( \sum_{j \neq i} M_{kj} (Y_j(0) - Y_k(0))\right)
\end{align*}
is unbiased for the middle term, using that $M_{kj} (Y_j(0) - Y_k(0)) = 0$ for $k=j$.
It follows that
\begin{align*}
    \hat{\mmv}_i
    &=
    \frac{1}{N{-}3} \sum_{k \neq i} \left(\sum_{j \neq i} M_{kj} (Y_j(0) - Y_k(0))\right)^2 - \frac{1}{(N{-}3)(N{-}2)} \sum_{k, j \neq i} M^2_{kj} (Y_j(0) - Y_k(0))^2
    \\
    &\phantom{=} +
    \frac{2}{N-2} \sum_{k \neq i} M_{k0} \left( \sum_{j \neq i} M_{kj} (Y_j(0) - Y_k(0))\right)
    +
    \frac{1}{N} \sum_{k} M^2_{k0}
\end{align*}
is an unbiased estimator of the conditional variance $\mmv$.
\end{proof}

\begin{proof}[Proof of Proposition 2]
We adopt the notation from the informal proof following the proposition.
Write $\mathbf{\beta}_t = \widehat{\mathbf{M}}_{\cdot\cdot t}$ for matrix of GSC the weights corresponding to a treatment period $t \geq N+1$.
We note that for every $t$ the GSC solution $\mathbf{\beta}_t$ does not depend on $T$, by our assumption that the constraints are not time-dependent and using that the optimization problem in equation 2.5 is only considering past observations.
The matrix $\mathbf{\beta}_t$ minimizes (over ${\cal B}$) the convex quadratic form
\[
    \sum_{i=1}^N \frac{1}{t-1} \sum_{s = 1}^{t-1} \left(\beta_{i0} + \sum_{j=1}^N \beta_{ij} Y_{js}(0)\right)^2
    =
    \sum_{i=1}^N
    \mathbf{\beta}_i'
    \left(\tilde{\mathbf{\Sigma}}_{1\cdots t} + \tilde{\mathbf{\mu}}_{1\cdots t} \tilde{\mathbf{\mu}}_{1\cdots t}'\right)
    \mathbf{\beta}_i
\]
where $\mathbf{\beta}_i$ is the $N+1$-vector of $\beta_{ij}$
and
we write
\begin{align*}
    \mathbf{y}_u
    &= \begin{psmallmatrix}
        1 \\
        Y_{1u}(0) \\
        \vdots \\
        Y_{Nu}(0)
    \end{psmallmatrix},
    &
    \tilde{\mathbf{\mu}}_{s \cdots t} 
    &=
    \frac{1}{t-s+1}
    \sum_{u=s}^t 
    \mathbf{y}_u,
    &
    \tilde{\mathbf{\Sigma}}_{s \cdots t} 
    &=
    \frac{1}{t-s+1}
    \sum_{u=s}^t 
    \left(
    \mathbf{y}_u
        {-}
        \tilde{\mathbf{\mu}}_{s \cdots t} 
    \right)
    \left(
    \mathbf{y}_u
        {-}
        \tilde{\mathbf{\mu}}_{s \cdots t} 
    \right)'.
\end{align*}
Similarly, $\mathbf{\beta}^*_T$ minimizes (over ${\cal B}$)
\[
    \sum_{i=1}^N \frac{1}{T-N-1} \sum_{t = N+2}^{T} \left(\beta_{i0} + \sum_{j=1}^N \beta_{ij} Y_{jt}(0)\right)^2
    =
    \sum_{i=1}^N
    \mathbf{\beta}_i'
    \left(\tilde{\mathbf{\Sigma}}_{N+2\cdots T} + \tilde{\mathbf{\mu}}_{N+2\cdots T} \tilde{\mathbf{\mu}}_{N+2\cdots T}'\right)
    \mathbf{\beta}_i
\]

By assumption 3,
$\tilde{\mathbf{\Sigma}}_{1\cdots t} + \tilde{\mathbf{\mu}}_{1\cdots t} \tilde{\mathbf{\mu}}_{1\cdots t}' \rightarrow \tilde{\mathbf{\Sigma}} + \tilde{\mathbf{\mu}} \tilde{\mathbf{\mu}}'$
as $t \rightarrow \infty$
and
$\tilde{\mathbf{\Sigma}}_{N+2\cdots T} + \tilde{\mathbf{\mu}}_{N+2\cdots T} \tilde{\mathbf{\mu}}_{N+2\cdots T}' \rightarrow \tilde{\mathbf{\Sigma}} + \tilde{\mathbf{\mu}} \tilde{\mathbf{\mu}}'$
as $T \rightarrow \infty$,
where $\tilde{\mathbf{\mu}} = \begin{psmallmatrix}
    1 \\
    \mathbf{\mu}
\end{psmallmatrix}, \tilde{\mathbf{\Sigma}} = \begin{psmallmatrix}
    0 & \mathbf{0}' \\
    \mathbf{0} & \mathbf{\Sigma}
\end{psmallmatrix}$ and $\tilde{\mathbf{\Sigma}} + \tilde{\mathbf{\mu}} \tilde{\mathbf{\mu}}'$ positive definite.
As a consequence, $\mathbf{\beta}_t$ (as $t\rightarrow \infty$) and $ \mathbf{\beta}^*_T$  (as $T\rightarrow \infty$) converge to the same limit
$
    \mathbf{\beta}^* = \argmin_{\mathbf{\beta} \in {\cal B}}
    \sum_{i=1}^N \mathbf{\beta}_i' (\tilde{\mathbf{\Sigma}} + \tilde{\mathbf{\mu}} \tilde{\mathbf{\mu}}') \mathbf{\beta}_i.
$
To complete the argument laid out in section 3.6 it remains to show that the expected loss $\mmv({\cal M})$ of $\mathbf{\beta}_t$ with $t$ randomly chosen among $\{N+2,\ldots,T\}$ converges to the same limit as the loss $\mmv^*({\cal M})$ of $\mathbf{\beta}^*_T$.
Specifically, for horizon $T$,
\begin{align*}
    &|\mmv({\cal M}) - \mmv^*({\cal M})| \leq
     \frac{1}{N}
     \sum_{i=1}^N 
     \left|
     \frac{1}{T-N-1}
     \sum_{t = N+2}^{T} \left(\mathbf{y}_t' \mathbf{\beta}_{ti}\right)^2 - \left(\mathbf{y}_t' \mathbf{\beta}^*_{Ti}\right)^2
    \right|
    \\
    &=\frac{1}{N}
     \sum_{i=1}^N 
     \left|
     \frac{1}{T-N-1}
     \sum_{t = N+2}^{T} \left(\mathbf{y}_t' (\mathbf{\beta}_{ti} - \mathbf{\beta}^*_{Ti})\right)^2 + 2 \left(\mathbf{y}_t' (\mathbf{\beta}_{ti} - \mathbf{\beta}^*_{Ti})\right) \mathbf{y}_t' \mathbf{\beta}^*_{Ti}
    \right|
    \\
    &\leq
    \frac{1}{N (T-N-1)}
     \sum_{i=1}^N  \left(
     \sum_{t = N+2}^{T} \left(\mathbf{y}_t' (\mathbf{\beta}_{ti} - \mathbf{\beta}^*_{Ti})\right)^2
     + 2
     \sqrt{
     \sum_{t = N+2}^{T} \left(\mathbf{y}_t' (\mathbf{\beta}_{ti} - \mathbf{\beta}^*_{Ti})\right)^2}
     \sqrt{
     \sum_{t = N+2}^{T} (\mathbf{y}_t' \mathbf{\beta}^*_{Ti})^2 } \right).
\end{align*}
We have that
$
    \frac{1}{T-N-1} \sum_{t = N+2}^{T} (\mathbf{y}_t' \mathbf{\beta}^*_{Ti})^2 \rightarrow (\mathbf{\beta}^*_i)' (\tilde{\mathbf{\Sigma}} + \tilde{\mathbf{\mu}} \tilde{\mathbf{\mu}}') \mathbf{\beta}^*_i
$
and for all fixed $R \leq T$ that
\begin{align*}
    &\frac{1}{T-N-1} \sum_{t = N+2}^{T} \left(\mathbf{y}_t' (\mathbf{\beta}_{ti} - \mathbf{\beta}^*_{Ti})\right)^2
    \leq \frac{1}{T-N-1} 
    \sum_{t = N+2}^{T} \|\mathbf{y}_t\|^2 \|\mathbf{\beta}_{ti} - \mathbf{\beta}^*_{Ti}\|^2
    \\
    &\leq 
    \frac{1}{T-N-1} \sum_{t = N+2}^{R} \|\mathbf{y}_t\|^2 \|\mathbf{\beta}_{ti} - \mathbf{\beta}^*_{Ti}\|^2
    +
    \sup_{t > R} \|\mathbf{\beta}_{ti} - \mathbf{\beta}^*_{Ti}\|^2 \frac{1}{T-N-1}  \sum_{t = R+1}^{T} \|\mathbf{y}_t\|^2
\end{align*}
For every $\varepsilon > 0$ we can now choose some $R$ such that $\sup_{t,s > R} \|\mathbf{\beta}_{ti} - \mathbf{\beta}^*_{si}\|^2 < \varepsilon$, so
\begin{align*}
    \limsup_{T \rightarrow \infty}
    \frac{1}{T-N-1} \sum_{t = N+2}^{T} \left(\mathbf{y}_t' (\mathbf{\beta}_{ti} - \mathbf{\beta}^*_{Ti})\right)^2
    \leq \varepsilon \: \textnormal{trace}(\tilde{\mathbf{\Sigma}} + \tilde{\mathbf{\mu}} \tilde{\mathbf{\mu}}'),
\end{align*}
and thus $|\mmv({\cal M}) - \mmv^*({\cal M})| \rightarrow 0$ for $T \rightarrow \infty$.
\end{proof}

\newcommand{\1}{\mathbf{1}}
\newcommand{\0}{\mathbf{0}}
\newcommand{\R}{\mathbb{R}}

\begin{proof}[Proof of Proposition 3]
The result immediately follows from
    \begin{align*}
    \mme\left[\left.\hat\tau-\tau\right|\bv\right]
    &=
    \sumt V_t \sumi p_i^{(t)} \left(M_{i0t} + \sumj M_{ijt} Y_{jt}(0)\right)
    \\
    &=
    \sumt V_t \left(\sumj p_j^{(t)} M_{j0t} + \sumi \yitn \sumj p_j^{(t)} M_{jit}  \right).\qedhere
\end{align*}
\end{proof}

\begin{proof}[Proof of Proposition 4]
    The existence is a consequence of the existence of non-negative eigenvectors in the associated network (which can be established e.g. by the Perron--Frobenius theorem).
    For a direct proof, 
    note that the weight matrix 
    $\mathbf{M} \in \R^{N \times N}$
    defined by
    $M_{ij} = M_{ijt}$, $i,j \in \{1,\ldots,N\}$,
    fulfills
    \begin{align}
    \label{eqn:conformal}
        M_{ij} &\begin{cases}
            = 1, & i=j, \\
            \leq 0, & i \neq j
        \end{cases},
        &
        \mathbf{M}^{(t)} \1 &= \0.
    \end{align}
    
    By \eqref{eqn:conformal} $\mathbf{M}^{(t)}$ is singular, so there exists some vector $q \in \R^N \setminus \{\0\}$ such that $(\mathbf{M}^{(t)})' q = \0$.
        Define $p \in [0,1]^N, p'\1=1$ by $p_i = |q_i| / \sum_j |q_j|$.
        If $q$ has only non-negative or only non-positive values, we are done.
        Assume now that $q$ has both positive and negative elements $q_i$, and denote by $I^-$ the indices corresponding to negative elements, and by $I^+$ the indices corresponding to positive elements.
        Then
        \begin{align*}
            0 = \sum_{i \in I^-} \underbrace{((\mathbf{M}^{(t)})'q)_i}_{=0}
            =
            \underbrace{\sum_{i \in I^-} \sum_{j \in I^-} M_{ji} q_j}_{
            =
            \sum_{j \in I^-} \underbrace{q_j}_{< 0} \underbrace{\sum_{i \in I^-} M_{ji}}_{\geq 0} \leq 0} + \sum_{i \in I^-} \sum_{j \in I^+} \overbrace{M_{ji}}^{\leq 0} \underbrace{q_j}_{> 0}
             \leq 0
        \end{align*}
        and thus $\sum_{i \in I^-} M_{ji} = 0$ for all $j \in I^-$ and $M_{ji} = 0$ for all $(i,j) \in I^- \times I^+$.
        Since $\sum_{i} M_{ji} = 0$ and $M_{ji} \leq 0$ whenever $i \neq j$, it follows from $\sum_{i \in I^-} M_{ji} = 0$ that $M_{ji} = 0$ for all $i \notin I^{-}$, $j \in I^-$.
        We therefore have that
        \begin{align*}
            &\left(\sum_j |q_j|\right) ((\mathbf{M}^{(t)})'p)_i =
            \left(\sum_j |q_j|\right) \left(
            \sum_{j \in I^-} M_{ji} p_j + \sum_{j \in I^+} M_{ij} p_j\right)
            \\
            &=
            \begin{cases}
                - \sum_{j \in I^-} M_{ji} q_j,
                & i \in I^-, \\
                \sum_{j \in I^+} M_{ji} q_j,
                & i \notin I^-
            \end{cases}
            =
            \begin{cases}
                - ((\mathbf{M}^{(t)})'q)_i,
                & i \in I^-, \\
                ((\mathbf{M}^{(t)})'q)_i,
                & i \notin I^-
            \end{cases}
            = 0
        \end{align*}
        and thus $(\mathbf{M}^{(t)})' p = \0$.
\end{proof}

\begin{proof}[Proof of Proposition 5]
    By Perron--Frobenius \citep[e.g.][]{Berman1994-ph},
    for $\mathbf{W}$ non-negative and irreducible, $\mathbf{W}$ has a unique (up to scaling) non-negative eigenvector (which is positive), and that eigenvector corresponds to the largest eigenvalue of $\mathbf{W}$.
    Since $p$ in proposition 4 is non-negative and an eigenvector with eigenvalue 1, is is the same as the eigenvector centrality of the network $\mathbf{W}$.
\end{proof}

\begin{proof}[Proof of the variance estimator in \autoref{prop:multiple}]
    This proof generalized the proof of proposition 1 above.
    Specifically,
    for $[N] = \{1,\ldots,N\}$,
    \begin{align}
        \sum_{\substack{k \subseteq [N];\\ |k| = N_T}}
        \sum_{\substack{i \subseteq [N] \setminus k; \\ |i| = N_T}}
        \sum_{j,j'  \in [N] \setminus k \cup \{0\}}
        \frac{1}{\binom{|[N] \setminus (k \cup[ \{j,j'\})| }{ N_T}}
        a_{i,j,j'}
        =
        \sum_{\substack{k \subseteq \{1,\ldots,N\};\\ |k| = N_T}}
        \sum_{j,j'\in [N] \cup \{0\}}
        a_{k,j,j'}
    \label{eq:counting}
    \end{align}
    for a conformal tensor $a$.
    
    For fixed $t$ as above consider weights $M_{k j}$ indexed by $k \subseteq [N]$ with $|k|=N_T$ and $j \in [N] \cup 0$, for which (a) $\sum_{j=1}^N M_{k j} = 0$ and (b) $M_{kj} = 1$ for $j \in k$, and potential outcomes $Y_j(0)$ with $j \in [N]$.
    Write $\overline{Y}_k(0) = \frac{1}{N_T} \sum_{j \in k} Y_j(0)$.
    (This approach generalizes to cases where treated units are themselves weighted, in which case we would replace $\overline{Y}_k(0)$ by the corresponding weighted average.)
    Let
    \begin{align}
        b_{k,j} &= 
        \begin{cases}
            0, 
            & j \in k,
            \\
            M_{kj} (Y_j(0) - \overline{Y}_k(0)), 
            & j \in [N] \setminus k,
            \\
            M_{k0},
            & j = 0,
        \end{cases}
        &
        a_{k,j,j'} &= b_{k,j} b_{k,j'}.
        \label{eq:bkj}
    \end{align}
    Then, for $K = \binom{N }{ N_T}$, and using that (c) $a_{k,j,j'} = 0$ whenever $j$ or $j'$ are in $[N] \setminus k$ and (d) $a_{k,j,j'} = a_{k,j',j}$,
    {\allowdisplaybreaks
    \begin{align*}
        \mmv
        &=
        \frac{1}{K}
        \sum_{\substack{k \subseteq [N];\\ |k| = N_T}}
        \left(
            M_{k0}
            +
            \sum_{j=1}^N M_{kj} Y_j(0)
        \right)^2
        \stackrel{\text{(a)}}{=}
        \frac{1}{K}
        \sum_{\substack{k \subseteq [N];\\ |k| = N_T}}
        \left(
            M_{k0}
            +
            \sum_{j=1}^N M_{kj} (Y_j(0) - \overline{Y}_k(0))
        \right)^2
        \\
        &\stackrel{\text{(b)}}{=}
        \frac{1}{K}
        \sum_{\substack{k \subseteq [N];\\ |k| = N_T}}
        \left(
            M_{k0}
            +
            \sum_{j \in [N] \setminus k} M_{kj} (Y_j(0) - \overline{Y}_k(0))
        \right)^2
        \stackrel{\text{(\ref{eq:bkj}.1)}}{=}
        \frac{1}{K}
        \sum_{\substack{k \subseteq [N];\\ |k| = N_T}}
        \left(
            \sum_{j=0}^N b_{kj}
        \right)^2
        \\
        &\stackrel{\text{(\ref{eq:bkj}.2)}}{=}
        \frac{1}{K}
        \sum_{\substack{k \subseteq [N];\\ |k| = N_T}}
            \sum_{j,j'\in [N] \cup \{0\}} a_{k,j,j'}
        \stackrel{\text{(\ref{eq:counting})}}{=}
        \frac{1}{K}
        \sum_{\substack{k \subseteq [N];\\ |k| = N_T}}
        \sum_{\substack{i \subseteq [N] \setminus k; \\ |i| = N_T}}
        \sum_{j,j'  \in [N] \setminus k \cup \{0\}}
        \frac{1}{\binom{|[N] \setminus (k \cup[ \{j,j'\})| }{ N_T}}
        a_{i,j,j'}
        \\
        & \stackrel{\text{(c)}}{=}
        \frac{1}{K}
        \sum_{\substack{k \subseteq [N];\\ |k| = N_T}}
        \sum_{\substack{i \subseteq [N] \setminus k; \\ |i| = N_T}}
        \Bigg(
            \frac{a_{i,0,0}}{\binom{N_C }{ N_T}}
            +
            \sum_{j  \in [N] \setminus k} \frac{
            a_{i,j,0}{+} a_{i,0,j} {+} a_{i,j,j}}{\binom{N_C - 1 }{ N_T}}
            +
            \sum_{\substack{j,j'  \in [N] \setminus k \\ j \neq j'}}
            \frac{a_{i,j,j'}}{\binom{N_C - 2 }{ N_T}}
        \Bigg)
        \\
        &\stackrel{\text{(d)}}{=}
        \frac{1}{K}
        \sum_{\substack{k \subseteq [N];\\ |k| = N_T}}
        \Bigg(
        a_{k,0,0}
        +
        \sum_{\substack{i \subseteq [N] \setminus k; \\ |i| = N_T}}
        \Bigg(
            \sum_{j  \in [N] \setminus k}
            \Big(
            \frac{2 a_{i,j,0}}{\binom{N_C - 1 }{ N_T}}
            -
            \underbrace{
            \frac{N_T}{(N_C - 1) \binom{N_C - 2 }{ N_T}}
            }_{
                = \frac{1}{\binom{N_C - 2 }{ N_T}}
            -
            \frac{1}{\binom{N_C - 1 }{ N_T}}
            }
            a_{i,j,j}
            \Big)
            +
            \sum_{\substack{j,j'  \in [N] \setminus k}}
            \frac{
            a_{i,j,j'}   }{\binom{N_C - 2 }{ N_T}}
        \Bigg)
        \Bigg)
        \\
        &\stackrel{\text{(\ref{eq:bkj}.2)}}{=}
        \frac{1}{K}
        \sum_{\substack{k \subseteq [N];\\ |k| = N_T}}
        \Bigg(
        b^2_{k,0}
        +
        \sum_{\substack{i \subseteq [N] \setminus k; \\ |i| = N_T}}
        \Bigg(
            \frac{2 b_{i,0} \sum_{j  \in [N] \setminus k} b_{i,j}}{\binom{N_C - 1 }{ N_T}}
            -
            \frac{N_T \sum_{j  \in [N] \setminus k} b^2_{i,j}}{(N_C - 1) \binom{N_C - 2 }{ N_T}}
            +
            \frac{
            \left(\sum_{j  \in [N] \setminus k} b_{i,j}\right)^2}{\binom{N_C - 2 }{ N_T}}
        \Bigg)
        \Bigg)
        \\
        &\stackrel{\text{(\ref{eq:bkj}.1)}}{=}
        \frac{1}{K}
        \Bigg(
        \frac{1}{K}
        \sum_{\substack{i \subseteq [N]; \\ |i| = N_T}}
        M_{i0}^2
        +
        \sum_{\substack{i \subseteq [N] \setminus k; \\ |i| = N_T}}
        \Bigg\{
            \frac{2}{\binom{N_C - 1 }{ N_T}}
            M_{i0} \sum_{j  \in [N] \setminus k} M_{ij} (Y_j - \overline{Y}_i)
        \\  
        &\hspace{3em}   
            +
            \frac{1}{\binom{N_C - 2 }{ N_T}}
            \left(\sum_{j \in [N] \setminus k}  M_{ij} (Y_j - \overline{Y}_i) \right)^2
            -
            \frac{N_T }{(N_C - 1) \binom{N_C - 2 }{ N_T}}
            \sum_{j  \in [N] \setminus k}
            M^2_{ij} (Y_j - \overline{Y}_i)^2
        \Bigg\}
        \Bigg)
        \\
        &= 
        \frac{1}{K}
        \sum_{\substack{k \subseteq [N];\\ |k| = N_T}}
        \hat{\mmv}_k,
    \end{align*}}
    so the proposed estimator is unbiased for the variance.
    
    An alternative estimator that emphasizes the leave-fold-out nature of this construction is
    \begin{align*}
        &
        \sum_{\substack{i \subseteq [N] \setminus k; \\ |i| = N_T}}
        \Bigg\{
        \frac{1}{\binom{N_C }{ N_T}} M_{i0}^2
        +
            \frac{2}{\binom{N_C - 1 }{ N_T}}
            M_{i0} \sum_{j  \in [N] \setminus k} M_{ij} (Y_j - \overline{Y}_i)
        \\  
        &\hspace{3em}   
            +
            \frac{1}{\binom{N_C - 2 }{ N_T}}
            \left(\sum_{j \in [N] \setminus k}  M_{ij} (Y_j - \overline{Y}_i) \right)^2
            -
            \frac{N_T }{(N_C - 1) \binom{N_C - 2 }{ N_T}}
            \sum_{j  \in [N] \setminus k}
            M^2_{ij} (Y_j - \overline{Y}_i)^2
        \Bigg\}.
    \end{align*}
    It has the additional advantage that it does not use the weights on the treated observations when constructing the variance for a specific draw.
    Here, the first average of squared intercepts could be replaced by the overall average of squared intercepts, as in the main variance estimator above.
\end{proof}

\section{GENERALIZATIONS AND EXTENSIONS}

In this section, we present three generalizations of the setting considered so far. The focus up to now was on the case with a single treated unit and single treated period where the estimand was the average effect for the treated under random assignment. First, we consider the case with multiple treated units. Second, we consider the case where the estimand is the average effect for all units in the treated period. Both of these generalizations create conceptual complications.

\subsection{Multiple Treated Units}
\label{section:multiple}

In this section, we look at the case with multiple treated units. We fix the number of treated units at $\nttt$. The estimand is, as before, the average effect for the $\nttt$ treated units.
We modify assumption 1 to
\begin{assumption}\label{ass_random_unit_a}{\sc (Random Assignment of Units)}
	\[ {\mathbb{P}}(\bu=\buu)=
	\left\{
	\begin{array}{ll}
	\left(\frac{ N!}{\nttt!\nccc!}
	\right)^{-1}
	\hskip1cm & \mathrm{if}\ \ u_i\in\{0,1\}\: \forall i,\ \  \sumi u_i=\nttt,\\
		0 & \mathrm{otherwise.}
	\end{array}\right. \]
\end{assumption}
In this case, it is convenient to work with the sets of units assigned to the treatment, rather than the individual units. 
There are $K=N!/(\nttt!\nccc!) = \binom{N }{ N_T}$ such sets. 
Of these, $(N-1)!/((\nttt-1)!\nccc!) = \binom{N-1 }{ N_T-1}$ include a given unit, such as unit 1, since there are $\binom{N-1 }{ N_T-1}$ combinations of the remaining units if that unit is treated.
This represents a fraction $\nttt/N$ of the total number of sets of $\nttt$ treated units. 

Let
$\tbu$ be the vector of length $K$ of indicators denoting which set of  $\nttt$ units is treated. 
Let $e_k$ be the $K$-component vector with the $k$-th component equal to one and  all other elements equal to zero. By construction, $\sum_{k=1}^{K}\tilde{U}_k=1$, and $\tilde{U}_k\in\{0,1\}$.
\autoref{ass_random_unit_a} implies that the probability that $\tilde{U}_k=1$ is equal to $1/K$.
Let $u_i(\tbu)\in\{0,1\}$ be an indicator for unit $i$ being treated given the assignment vector $\tbu$.
In this notation, we can rewrite $\tau$ as
\[ \tau=\tau(\tbu,\bv)=\frac{1}{\nttt} \sum_{k=1}^K\sumt \tilde{U}_k V_t
\sum_{i=1}^N u_i(\tilde{U}_k)\Bigl(\yite-\yitn\Bigr).\]

Instead of the tensors $\momega$ with dimension $N\times (N+1)\times T$, we now have tensors with dimension $K\times (N+1)\times T$, with one row for each of the  $K=N!/(\nttt!\nccc!)$ possible sets of treated units.
The estimators we consider are of the form
\begin{equation}
    \hat\tau(\tbu,\bv,\by,\momega)\equiv
    \sum_{k=1}^K	\sumt  
    \tilde{U}_k
    V_t \left\{M_{k0t}+ \sumj  M_{kjt} Y_{jt}\right\}.
\end{equation}
This formulation suggests the restriction that
$
M_{kjt}=1/N_T,
$
for all $j$ such that $u_j(e_k)=1$ and $
M_{kjt}\leq0,
$ whenever $u_j(e_k)=0$.
The set of such $\momega$ we  consider for the generalized modified unbiased Synthetic Control (MUSC) estimator is
\[ \calm^\musc=\left\{\momega\middle| \sum_{j=1}^{N}M_{kjt}=0\: \forall k,t, \sum_{k=1}^{K}M_{kjt}=0\: \forall j\geq 1,t\right\}.\]
The objective function for choosing $\momega$ is now
\[\momega(\by,\calm^\musc)=\argmin_{\momega\in\calm^\musc} \sum_{k=1}^K \sumt\left\{\sum_{s< t}\left(M_{k0t}+\sumj M_{kjt} Y_{js}\right)^2\right\}.\]

\begin{prop}
\label{prop:multiple}
Suppose that \autoref{ass_random_unit_a} holds. Then\\
$(i)$ the estimator $\hat\tau(\tbu,\bv,\by,\momega(\by,\calm^\musc))$ is unbiased conditional on $\bv$,
$(ii)$ the variance of  $\hat\tau(\tbu,\bv,\by,\momega(\by,\calm^\musc)$ is
\[
    \mmv\left(\hat\tau(\tbu,\bv,\by,\momega(\by,\calm^\musc))\middle|\bv\right) =
    \frac{1}{K} \sum_{t=1}^T V_t \sum_{k=1}^K \left(M_{k0t}+\sumj M_{kjt} Y_{ts}(0)\right)^2,    
\]
and
$(iii)$, the variance can be estimated without bias (conditional on $\bv$) by a generalization of the variance estimator in proposition 1,
\begin{align*}
    \hat{\mmv}
    =
    \sum_{k=1}^K\sumt \tilde{U}_k & V_t \Bigg(
    \sum_{\substack{k'=1; \\ \mathclap{u_i(\tilde{U}_k) + u_i(\tilde{U}_{k'}) \leq 1 \forall i}}}^K
    \Bigg\{
        \frac{1}{\binom{N_C - 2 }{ N_T}}
        \left(\sum_{j=1}^N  (1-u_j(\tilde{U}_k)) M_{k'jt} (Y_{jt} - \overline{Y}_{k't}) \right)^2
        \\
        &-
        \frac{N_T }{(N_C - 1) \binom{N_C - 2 }{ N_T}}
        \sum_{j=1}^N (1-u_j(\tilde{U}_k))
        M_{k'jt}^2 (Y_{jt} - \overline{Y}_{k't})^2
        \\  
        &   
        + 
        \frac{2}{\binom{N_C - 1 }{ N_T}}
        M_{k'0t} \sum_{j=1}^N (1-u_j(\tilde{U}_k)) M_{k'jt} (Y_j - \overline{Y}_{k'})
        \Bigg\}
        +\frac{1}{K}
    \sum_{k'=1}^K
    M_{k'0t}^2 \Bigg)
\end{align*}
for $\overline{Y}_{k't} = \frac{1}{N_T} \sum_{j=1}^N u_j(\tilde{U}_{k'}) Y_{jt}$.
\end{prop}

\subsection{The Average Effect for All Units}\label{all_treated}

Here we look at the case  where the estimand changes from the average effect for the treated unit(s) to the average effect over all units in the treated periods. For ease of exposition, we continue to focus on the case with a single treated period and a single treated unit. Formally, the estimand is
\[ \tau^\vvv=\tau^\vvv(\bv)\equiv\frac{1}{N} \sumi\sumt V_t \Bigl(\yite-\yitn\Bigr),\]
We can separate this into two components, the  effect for the treated unit, and the average effect for the controls
\[ \taut\equiv\sumi\sumt U_iV_t \Bigl(\yite-\yitn\Bigr),
\ \ \ 
\tau^\ccc\equiv\frac{1}{N-1}\sumi\sumt (1-U_i)V_t \Bigl(\yite-\yitn\Bigr),
\]
with $\tau^\vvv=\taut/N+\tau^\ccc(N-1)/N$.
Consider, as before, an estimator of the form
\[\hat\tau(\bu,\bv,\by,\momega)=
\sumituv\left\{M_{i0t}+ \sum_{j=1}^N  M_{ijt} Y_{jt}\right\}.
\]
 The restrictions $M_{iit} = 1 \: \forall i,t$, $\sum_{i=1}^N M_{ijt} = 0 \: \forall j, t$ (including the intercept) still imply unbiasedness conditional on $\bv$, and the MUSC remains unbiased for $\tau^V$.
 Yet the variance (and more generally the conditional expected loss of such a weighted estimator) is now
 \begin{align}
     &\mme
 \left[ \left(\hat\tau(\bu,\bv,\by,\momega)-\tau^V\right)^2 \middle| \bv\right]=\frac{1}{N}\sumit V_t \bigg(
 M_{i0t}+ \left(M_{iit} - \frac{1}{N}\right) \yite
 \nonumber
 \\
 & \ \ \  -
 \sum_{j=1}^N (1-U_j) \frac{1}{N} Y_{jt}(1)+
 \frac{1}{N} Y_{it}(0)
 +
 \sum_{j=1}^N (1-U_j) \left(M_{ijt} + \frac{1}{N}\right) Y_{jt}(0) 
 \bigg)^2
 \label{eqn:ateerror}
 \end{align}
which depends on treated and untreated potential outcomes.
This creates two challenges.
First, since the expression depends on treated outcomes, there is no immediate sample analogue that corresponds to minimizing expected error, even under time randomization.
Second, the variance cannot generally be estimated without bias, since it depends not only on the variation of the $Y_{it}(0)$ (which can be estimated), but also on the variation of the $Y_{it}(1)$ and their covariance with the $Y_{it}(0)$ (neither of which is identified from the data).

We briefly discuss two possibilities 
for the (non-stochastic) correlation of treatment and control outcomes, and what they imply for estimation.
First, if treatment effects are constant within time period (and treatment and control potential outcomes thus perfectly correlated, $Y_{it}(1) - Y_{it}(0) = \tau$), then (\ref{eqn:ateerror}) becomes
\begin{align*}
    \frac{1}{N}\sumit V_t \bigg(
     M_{i0t}+
     \sum_{j=1}^N M_{ijt} Y_{jt}(0) 
     \bigg)^2
\end{align*}
as before, suggesting the MUSC estimator.
{If, on the other hand, treated outcomes are uncorrelated to control outcomes,
then \eqref{eqn:ateerror} becomes \begin{align*}\frac{1}{N}\sumit V_t \bigg(     M_{i0t}      +      \frac{1}{N} \sum_{j=1}^N Y_{jt}(0)      +      \sum_{j=1}^N (1-U_j) M_{ijt} Y_{jt}(0)       \bigg)^2      + \text{const.},
\end{align*}
which suggests an alternative MUSC-type estimator that minimizes the sample analogue in non-treated time periods over weights $\calm^\musc$, which could effectively shrink the MUSC weights on control units towards the DiM weights.}

\end{document}